\documentclass{aa}

\usepackage{array}

\newcolumntype{M}[1]{>{\centering\arraybackslash}m{#1}}
\newcolumntype{N}{@{}m{0pt}@{}}

\usepackage{rotating}
\usepackage{graphicx}
\usepackage{graphics}
\usepackage{amssymb}
\usepackage{color}
\usepackage{natbib}
\bibpunct{(}{)}{;}{a}{}{,}
\def\Len{L_{\rm en}}
\def\Lhy{L_{\rm hy}}
\def\Reff{R_{\rm eff}}
\def\Fxuv{F_{\rm XUV}}

\def\qxuv{Q_{\rm XUV}}
\def\La{Q_{\rm Ly-\alpha}}
\def\hhh{\rm H_3^+}
\def\qhhh{Q_{\hhh}}
\def\hh{\rm H_2}
\def\hp{\rm H^+}
\def\hhp{\hh^+}
\def\roche{R_{\rm roche}}
\def\nh{n_{\rm H}}
\def\nhh{n_{\hh}}
\def\nhhh{n_{\hhh}}
\def\nhp{n_{\hp}}
\def\nhhp{n_{\hhp}}
\def\ne{n_{\rm e}}
\def\ergscm{\rm erg\,s^{-1}\,cm^{-2}}
\def\Teq{$T_{\rm eq}$}

\begin{document}

 \title{Young planets under extreme UV irradiation. I. Upper atmosphere modelling of the young exoplanet K2-33b}
  \author{D.~Kubyshkina\inst{1}
          \and
          M.~Lendl\inst{1,2}
          \and
          L.~Fossati\inst{1}
          \and
          P.~E.~Cubillos\inst{1}
          \and
          H.~Lammer\inst{1}
          \and
          N.~V.~Erkaev\inst{3,4}
          \and
          C.~P.~Johnstone\inst{5}
          }
\institute{
        Space Research Institute, Austrian Academy of Sciences, Schmiedlstr. 6,         8042 Graz, Austria\\
        \email{daria.kubyshkina@oeaw.ac.at}
        \and
        Max Planck Institute for Astronomy, K\"onigstuhl 17, 69117 Heidelberg,          Germany
        \and
    Institute of Computational Modelling, FRC ``Krasnoyarsk Science Center'' SB RAS,``'', 660036, Krasnoyarsk, Russian Federation
        \and
    Siberian Federal University, 660041, Krasnoyarsk, Russian Federation
        \and
    University of Vienna, Department of Astrophysics, T\"urkenschanzstrasse         17, 1180, Vienna, Austria
}

\date{}
\abstract
{The K2-33 planetary system hosts one transiting
$\sim$5\,$R_{\oplus}$ planet orbiting the young M-type host
star. The planet's mass is still unknown, with an estimated upper
limit of $5.4$~$M_J$. The extreme youth of the system ($<$20\,Myr) gives the unprecedented opportunity to study the
earliest phases of planetary evolution, at a stage when the
planet is exposed to an extremely high level of high-energy radiation
emitted by the host star. We perform a series of 1D hydrodynamic
simulations of the planet's upper atmosphere considering a range
of possible planetary masses, from 2 to 40 $M_{\oplus}$, and
equilibrium temperatures, from 850 to 1300\,K, to account for
internal heating as a result of contraction. We obtain temperature
profiles mostly controlled by the planet's mass, while the
equilibrium temperature has a secondary effect. For planetary
masses below 7--10\,$M_{\oplus}$, the atmosphere is
subject to extremely high escape rates, driven by the planet's
weak gravity and high thermal energy, which increase with
decreasing mass and/or increasing temperature. For higher masses,
the escape is instead driven by the absorption of the high-energy
stellar radiation. A rough comparison of the timescales for complete
atmospheric escape and age of the system indicates that the planet
is more massive than 10\,$M_{\oplus}$.}
\titlerunning{Young planets under extreme UV irradiation}
\authorrunning{D. Kubyshkina et al.}
\maketitle
%

%
%
\section{Introduction}\label{sec:intro}
With the number of extrasolar planets that we have discovered
rapidly increasing and the development of characterization
techniques, the comparative analysis of exoplanet properties is
becoming possible. To date, this effort has focused largely on the
comparison of planets of different masses and stellar irradiation
levels. The temporal component of this complex question, that is,
the evolution of planetary atmospheres with time, has been touched
upon mostly in the context of planets orbiting Gyrs old stars,
evolved away from the zero-age main sequence, and the reaction of
planetary atmospheres to slowly increasing irradiation caused by a
host star evolving off the main sequence
\citep[e.g.][]{owen2016a,lopez2016}. This is mostly because the
early evolution of planetary atmospheres, which is the most
relevant in terms of atmospheric evolution
\citep[e.g.][]{owen2013,lopez2013,jin2014,chen2016}, has been
inaccessible due to various observational difficulties related to
the activity and variability of young stars that have precluded
the detection of planets in their early phases of evolution.

After the dispersal of the protoplanetary disk, extrasolar planets
cool and contract to their currently observed sizes
\citep[e.g.][]{baraffe2006,Steokl2016}. {Recently,
\citet{stoekl2015} and }\citet{owen2016b} suggested that planets
emerging from their disks rapidly find themselves in a state of
strong mass loss, called ``boil-off'', due to Parker winds in the
planetary atmosphere driven by the stellar thermal irradiation.
During this short lived phase, the planet rapidly cools and
contracts, losing a substantial fraction of its gaseous envelope
before the planetary atmosphere settles into a more stable
configuration and continues to evolve on slower timescales
\citep[e.g.][]{jin2014,Fossati17,jin2017,owen2017}. For planets
close enough to the host star, the atmosphere may then go through
a long phase of efficient hydrodynamic escape, or even
``blow-off'', which, depending on the the planet's atmosphere
mass, may or may not significantly affect the planet
\citep[e.g.][]{lammer2003,yelle2004,lecavelier2004}.

In this work, we present simulations of the upper atmospheric
structure of the youngest confirmed exoplanet, K2-33b. Identified
from $K2$ photometry by \citet{Vanderburg16} and determined to be
a pre-main-sequence low-mass member of the Upper Scorpius OB
association by \citet{david2016} and \citet{mann2016}, K2-33b is a
planet with a radius of about 5\,$R_{\oplus}$ (i.e. 1.3 times
Neptune's radius). {The semi-major axis is 0.0409\,au and the
rotational period is 5.43 days.} The planetary mass is
undetermined, yet a conservative upper limit of 5.4\,$M_{\rm J}$
\citep{mann2016} indicates that the object is clearly a planet.
Owing to the planet's size, the true mass is likely to be somewhat
smaller than that of Jupiter (i.e. $\approx$318\,$M_{\oplus}$).
The K2-33 system still possesses a protoplanetary disk; however,
the lack of emission from hot material indicates that the disk
appears to be truncated at 2 AU \citep{david2016}. Age estimates
for the K2-33 system are 5--10\,Myr \citep{david2016} and
$<20$\,Myr \citep{mann2016}.

The planet K2-33b is thus an object that has only very recently emerged from
its birthplace inside the protoplanetary disk in which it was
formed, and its atmosphere may still be in the process of
cooling and contracting to a more stable configuration. The planet
may either have formed further out in the system and
travelled to its present location via disk-driven migration
\citep[e.g.][]{lin1996,ida2008,rafikov2006,schlaufman2009,lubow2010,raymond2014},
or have formed at (or very near) its present location
\citep{chiang2013,ogihara2015}. Owing to the young age of the
system, the planet is likely experiencing extremely high levels of
high-energy X-ray and extreme ultraviolet (XUV) stellar
irradiation \citep[e.g.][]{wright2011}. Its youth makes this
object a key target for atmospheric studies, providing a snapshot
on the earliest stages of planetary evolution.

We present here a grid of upper atmosphere models for K2-33b,
aiming to determine the possible state of this planet's atmosphere
and to look for possible constraints on the planet's mass from
upper atmosphere modelling. In Section~\ref{sec:model} we present
the physical model used in our calculations, the results of which
are presented in Section~\ref{sec:results}. In
Section~\ref{sec:discussion} we offer a first interpretation of
the possible atmospheric properties of K2-33b and summarize our
conclusions in Section~\ref{sect:conclusions}.

\section{Modelling setup and formalism }\label{sec:model}

\subsection{System parameters}\label{sec:param}

\citet{david2016} and \citet{mann2016} estimate compatible values
of 1.1$\pm$0.1 and 1.05$\pm$0.07\,$R_{\odot}$ for the radius of
the host star K2-33. From the rather well determined stellar
radius follows a planetary radius of
5.76$^{+0.62}_{-0.58}$\,$R_{\oplus}$ \citep{david2016}, or
5.04$^{+0.34}_{-0.37}$\,$R_{\oplus}$ \citep{mann2016}. However,
stellar mass estimates diverge depending on whether or not the
stellar density, inferred from the transit light curve
\citep[following][]{Seager03}, is included in the analysis. By
considering stellar densities inferred from the transit light
curve (0.49$^{+0.21}_{-0.27}$\,$\rho_{\odot}$), \citet{mann2016}
find a stellar mass of 0.56$\pm$0.09\,$M_{\odot}$, while
\citet{david2016} report a mass of 0.31$\pm$0.05\,$M_{\odot}$ and
a corresponding stellar density of 0.34$\pm$0.12\,$\rho_{\odot}$,
based on stellar evolutionary models. For our calculations, we
base ourselves on \citet{mann2016} and use therefore a value of
0.56$\pm$0.09\,$M_{\odot}$. We also consider an orbital
separation of $d$\,=\,0.04\,au.

We estimated the stellar X-ray luminosity from the basic stellar
parameters: mass, radius, and rotation rate. The X-ray luminosity, $L_{\rm
X}$ , can be calculated from \citep[e.g.][]{wright2011}
\begin{equation*}
R_{\rm X} = \frac{L_{\rm X}}{L_{\rm bol}} =
    \begin{cases}
    R_{\rm X,sat} &\text{if $R_0<R_{\rm 0,sat}$}\\
    CR_0^{\beta} &\text{if $R_0\geq R_{\rm 0,sat}$}\,,
    \end{cases}
\end{equation*}
where $R_{\rm 0,sat}$ is the saturation threshold corresponding to
the Rossby number $R_0$, which is the ratio between the stellar
rotation period and the convective turnover time ($R_{\rm 0,sat} =
0.13$; \citealt{wright2011}). The X-ray emission reaches the
saturation value $R_{\rm X,sat}$ for the fast rotating stars, with
$R_0$ smaller than $R_{\rm 0,sat}$. \citet{mann2016} estimated the
stellar rotation period to be 6.27$\pm$0.17\,days and the
theoretical prediction of the convection turnover time for a
stellar mass of 0.5\,$M_{\odot}$ is 35\,days \citep{wright2011}.
This implies that the star is already out of the saturated regime.
By using values of $\beta$\,=\,$-$2.18 and
$C$\,=\,8.68$\times10^{-6}$ \citep{pizzolato2003,wright2011}, we
obtain $L_{\rm X}$\,=\,2.16$\times10^{29}$\,erg\,s$^{-1}$ and
consequently an X-ray flux at the planetary orbit of
48,257\,$\ergscm$, which is about 10$^5$ times stronger than that
currently experienced at Earth.

The X-ray and  extreme ultraviolet (EUV) luminosities are related
as \citep{sf2011}
\begin{equation}
\log{L_{\rm EUV}} = (4.80 \pm 1.99) + (0.860 \pm 0.073)\log{L_{\rm X}}\,,
\end{equation}
which leads to a stellar EUV luminosity of $L_{\rm
EUV}$\,=\,1.066$\times10^{30}$\,erg\,s$^{-1}$ and an EUV flux at
the planet's orbit of 2.382$\times10^{5}$\,$\ergscm$.

Since the planetary mass is essentially unknown, we perform our
simulations over a grid in planetary mass, assuming values of 2,
4, 7, 10, 20, and 40\,$M_{\oplus}$. We also consider different
values of the equilibrium temperature (\Teq) of 850, 1000, 1150,
and 1300\,K, where the lowest value of 850\,K is roughly the \Teq\
value calculated on the basis of the system parameters
\citep{mann2016}, assuming a Bond albedo of zero and zero energy
redistribution. By considering higher temperatures, we account for
possible additional internal heating originating from gravitational
contraction and radioactive decay.

\subsection{Physical model}\label{sec:physical_model}
To model the planetary upper atmosphere and infer the mass-loss
rates, we employ an updated version of the 1D hydrodynamical code
described in detail in the appendix of \citet{erkaev2016}. This
model treats the planetary upper atmosphere as a clear hydrogen
gas envelope, including the effects of recombination,
dissociation, ionization of the hydrogen atoms and molecules, and
Ly-$\alpha$ cooling. In this work, we additionally include cooling
from $\hhh$ molecules (see details below).

The original code does not account for the spectral dependence {of
stellar} high-energy flux and considers just the integrated EUV
flux, excluding X-rays. To make the model more appropriate for
treating young stars with intense X-ray fluxes, such as K2-33, we
implemented also X-ray heating (see details below).

The lower boundary of the calculation domain, $R_{\rm pl}$, is set
to be at the planetary photospheric radius, where the optical and
infrared stellar photons are absorbed. The upper boundary is set
at the Roche radius $\roche = d\times(\frac{\alpha}{3})^{1/3}$,
where $\alpha = \frac{M_{\rm pl}}{M_{\rm pl}+M_*}$, and $M_{\rm
pl}$ and $M_*$ are the planetary and stellar masses, respectively.
We also assume that the atmosphere at the lower boundary consists
exclusively of molecular hydrogen, while the rest of the
atmosphere is composed of molecular and atomic hydrogen, their
ions, and $\hhh$.

For each modelled planet, the atmospheric pressure at the
planetary photospheric radius (i.e. the lower boundary) was
estimated employing the same technique described in
\citet{Fossati17} and \citet{cubillos2017}, which is based on the
radiative transfer code BART \citep[Bayesian Atmospheric Radiative
Transfer,][]{bart}, and assuming solar metallicity. The obtained
pressure values, which range between about 40 and 400\,mbar, are
listed in Table~1. We further set the temperature at the lower
boundary to be equal to the equilibrium temperature
\citep[see][for a discussion on this point]{Fossati17}.
\begin{table}[h]
\label{tab:tab1}
\caption{Dependence of the lower pressure boundary (in mbar) as a function of planetary \Teq\ and mass ($M_{\rm pl}$).}
\begin{tabular}{c|c|c|c|c}
  \hline
  $M_{\rm pl}$ [$M_{\oplus}$] $\diagdown$ \Teq\ [K] & 850 & 1000 & 1150 & 1300 \\
  \hline
  2.0 & 40.8 & 39.5 & 39.1 & 39.0 \\
  4.0 & 73.4 & 70.4 & 67.4 & 65.0 \\
  7.0 & 112.7 & 108.9 & 105.3 & 103.1 \\
  10.0 & 145.5 & 140.5 & 135.8 & 133.4 \\
  15.0 & 192.6 & 185.5 & 178.8 & 176.0 \\
  20.0 & 233.8 & 224.7 & 216.1 & 212.8 \\
  40.0 & 368.9 & 351.9 & 336.0 & 330.7 \\
  \hline
\end{tabular}
\end{table}

The height-averaged heating efficiency ($\eta$) of the upper
atmosphere exposed to high-energy stellar radiation is usually
considered to be between 10\%\ and 60\%
\citep[e.g.][]{Watson1981,yelle2004,mc2009,cp2009,owen2012,shematovich2014,Salz2016}.
In this work, we follow the considerations of \citet{erkaev2016}
and adopt a value of 15\%, based on the results of the direct
simulation Monte Carlo model calculations of Shematovich et al.
(2014), which model photolytic and electron impact processes in the
thermosphere by solving the kinetic Boltzmann equation.

With the inclusion of Ly-$\alpha$ and $\hhh-$cooling, the equations for mass, momentum, and energy conservation are respectively
\begin{eqnarray}
\frac{\partial\rho}{\partial t} + \frac{\partial(\rho v
r^2)}{r^2\partial r} &=& 0\,, \\
\frac{\partial\rho v}{\partial t} + \frac{\partial[r^2(\rho
v^2+P)]}{r^2\partial r} &=& - \frac{\partial U}{\partial r} + \frac{2P}{r}\,, \\
\frac{\partial[\frac{1}{2}\rho v^2+E+\rho U]}{\partial t} &+&
\frac{\partial vr^2[\frac{1}{2}\rho v^2+E+P+\rho U]}{r^2\partial
r} = \nonumber \\
\qxuv - \La &+& \frac{\partial}{r^2\partial r}(r^2\chi
\frac{\partial T}{\partial r}) - \qhhh.
\end{eqnarray}
Here $\rho$, $v$, $T$, and $P$ {are mass} density, velocity,
temperature, and pressure of the gas as a function of the radial
distance $r$ from $R_{\rm pl}$, respectively.\footnote{We note
that here $r$ and $v$ are called $R$ and $V$ in
\citet{erkaev2016}.} The thermal energy $E$ is equal to
$[\frac{3}{2}(\nh+\nhp+n_e) + \frac{5}{2}(\nhh+\nhhp)+3\nhhh]kT$.
The thermal conductivity is $\mathbf{\chi}$ , which is given by
$\chi = 4.45\times10^4(\frac{T}{1000})^{0.7}$ \citep{Watson1981},
{where the exponent 0.7 was chosen for the neutral gas
\citep{banks1973}}. The terms $\qxuv$ and $\La$ are the stellar
XUV volume heating rate and $Ly-\alpha$ cooling, respectively,
while $U = \frac{G M_{\rm pl}}{R_{\rm pl}}(1 - \frac{R_{\rm
pl}}{r})$ is the gravitational potential.

In the absence of a measured stellar XUV spectrum and to speed up
computations for the XUV stellar flux absorption, we follow the
approach of \citet{erkaev2016}, {who considered} that the whole
stellar EUV flux is emitted at a single wavelength of 60\,nm,
which corresponds to the position of the peak of the stellar EUV
emission. To account for X-ray heating, we additionally {consider
absorption of stellar X-ray photons} by the planetary atmosphere.
{Here, we assume} that they are all emitted at a wavelength of
5\,nm, which lies roughly in the middle of the X-ray wavelength
range. {To summarize, we set the integrated X-ray and EUV stellar
flux to be emitted at 5 and 60\,nm, respectively.} The XUV heating
function $\qxuv$ is therefore composed of two terms, $Q_{\rm EUV}$
and $Q_{\rm X}$, {which describe the heating by the EUV and X-ray
stellar flux, respectively. These two functions are constructed in
the same way, but the absorption cross-sections and absorption
functions of the stellar flux inside the planetary atmosphere are
defined at the two wavelengths given above. This results in a
redistribution of the energy absorption inside the atmosphere,
which, however, does not significantly affect the large-scale
characteristics of the atmosphere.}

{The total heating function is} $\qxuv$\,=\,$Q_{\rm EUV} + Q_{\rm
X}$. Each heating function is given by
\begin{equation}
Q_{m} = \eta\sigma_{m}(\nh+\nhh)\phi_{m},
\end{equation}
where $m$ stands for either EUV or X, $\sigma_{m}$ is the
absorption cross-section for the specific wavelength, and
$\phi_{m}$ is the flux absorption function, given by
\begin{equation}
\phi_{m} =
\frac{1}{4\pi}\int_{0}^{\frac{\pi}{2}+arccos(\frac{1}{r})}\{
J_{m}(r,\theta) \, 2\pi \, sin(\theta) \}d\theta.
\end{equation}
Here, $J_{m}(r,\theta)$ is {a} function {in} spherical coordinates
describing the spatial variation of the EUV, or X-ray, flux due to
atmospheric absorption \citep{erkaev2015} and $r$ in this case is
radial distance from the planetary centre, in units of $R_{\rm
pl}$.

Following \citet{mc2009}, the absorption cross-sections are given
by $\sigma = \sigma_0(\frac{E_{\lambda}}{E_{i}})^{-3}$, where
$E_{i}$ is the hydrogen ionization energy and $E_{\lambda}$ is the
photon energy at a specific wavelength range, that is $E_{\lambda}
= 20$\,eV in the EUV and 248\,eV in the X-ray domain. This gives
EUV flux absorption cross-sections of $2\times10^{-18}$\,cm$^{-3}$
and $1.2\times10^{-18}$\,cm$^{-3}$ for atomic and molecular
hydrogen, respectively. The X-ray absorption cross-section is
$\sigma_x \sim 0.5\times10^{-3} \sigma_{euv}$.

We calculate Ly-$\alpha$ cooling using \citep{Watson1981}
\begin{equation}
\La = 7.5\times10^{-19} \ne\nh e^{\left(-\frac{118348}{T}\right)}.
\end{equation}
We also include $\hhh$ cooling, as given by \citet{miller2013}, in the form of
\begin{equation}\label{eqn:h3c}
\qhhh = 4\pi \nhhh e^{\sum_{n} C_{n}T^n{}}\,,
\end{equation}
where $C_n$ are the temperature-dependent coefficients listed in
Table~5 of \citet{miller2013}. To include $\hhh$ in the model, we
add the following additional chemical reactions \citep[with rate
coefficients taken from][]{yelle2004}
\begin{equation}
\hhp + \hh \rightarrow \hhh + H, \hspace{0.5 cm} \hhh + H
\rightarrow \hhp + \hh \nonumber\,,
\end{equation}
both having a cross-section of $\gamma_{\hh} = 2\times10^{-9}$,
\begin{equation}
\hhh + e \rightarrow \hh + H \nonumber\,,
\end{equation}
with a cross-section of $\alpha_{\hhh1} = 2.9\times10^{-8}(\frac{300}{T_e})^{0.65}$, and
\begin{equation}
\hhh + e \rightarrow H + H + H \nonumber\,,
\end{equation}
with a cross-section of $\alpha_{\hhh2} =
8.6\times10^{-8}(\frac{300}{T_e})^{0.65}$. By including these
reactions, the {continuity} equations for the atmospheric species
become
\begin{eqnarray}\label{eqn:species}
\frac{\partial \nh}{\partial t}+\frac{\partial(\nh
vr^2)}{r^2\partial r} &=& -\nu_{H}\nh -
\nu_{Hcol}\ne\nh  \nonumber \\
+ \alpha_{H}\ne\nh &+& 2\alpha_{\hh}\ne\nhh +
2\nu_{dis}\nhh n \nonumber \\
- 2\gamma_{H}n\nh^2 &+& \gamma_{\hh}(\nhh\nhhp-\nh\nhhh) \nonumber \\
+ (\alpha_{\hhh1} &+& 3\alpha_{\hhh2})\nhhh\ne\,, \\
\frac{\partial \nhh}{\partial t}+\frac{\partial(\nhh
vr^2)}{r^2\partial r} &=&
-\nu_{\hh}\nhh - \nu_{dis}\nhh n \nonumber \\
+ \gamma_{H}n\nh^2 &+&
\gamma_{\hh}(\nh\nhhh - \nhh\nhhp) \nonumber \\
&+& \alpha_{\hhh1}\nhhh\ne\,, \\
\frac{\partial \nhp}{\partial t}+\frac{\partial(\nhp
vr^2)}{r^2\partial r} &=& \nu_{H}\nh +
\nu_{Hcol}\ne\nh \nonumber \\
&-& \alpha_{H}\ne\nhp\,, \\
\frac{\partial \nhhp}{\partial t}+\frac{\partial(\nhhp
vr^2)}{r^2\partial r} &=&
\nu_{\hh}\nhh - \alpha_{\hh}\ne\nhhp \nonumber \\
+
\gamma_{\hh}(\nh\nhhh &-& \nhh\nhhp)\,, \\
\frac{\partial \nhhh}{\partial t}+\frac{\partial(\nhhh
vr^2)}{r^2\partial r} &=&
\gamma_{\hh}(\nhh\nhhp-\nh\nhhh) \nonumber \\
- (\alpha_{\hhh1}&+&\alpha_{\hhh2})\nhhh\ne\,.
\end{eqnarray}
Assuming quasi-neutrality, the electron density is given by
\begin{equation}\label{eqn:ne}
    \ne = \nhp + \nhhp + \nhhh\,,
\end{equation}
while the total hydrogen number density is the sum of the number
densities of all species. The ionization and dissociation
reactions used in the above equations {and} their cross-sections
are listed in Table~2. {We note that our photo-electric
cross-sections do not account for X-ray ionization \citep[see for
example ][ for a more thorough approach]{lorenzani2001,locci2018},
which is, however, orders of magnitude smaller than that caused by
EUV irradiation.}

The atmospheric mass density $\rho$ and pressure $P$ are given by
\begin{equation}\label{eqn:rho}
    \rho = m_{\rm H}(\nh+n_{\hp}) + m_{\hh}(\nhh+\nhhp) +
    m_{\rm H_3}\nhhh
\end{equation}
and
\begin{equation}\label{eqn:Press0}
    P = (\nh + n_{\hp} + \nhh + \nhhp + \nhhh+\ne)kT \,,
\end{equation}
while normalization and numerical realization are the same as in \citet{erkaev2016}.
\begin{figure}[h!]
  \includegraphics[width=\hsize]{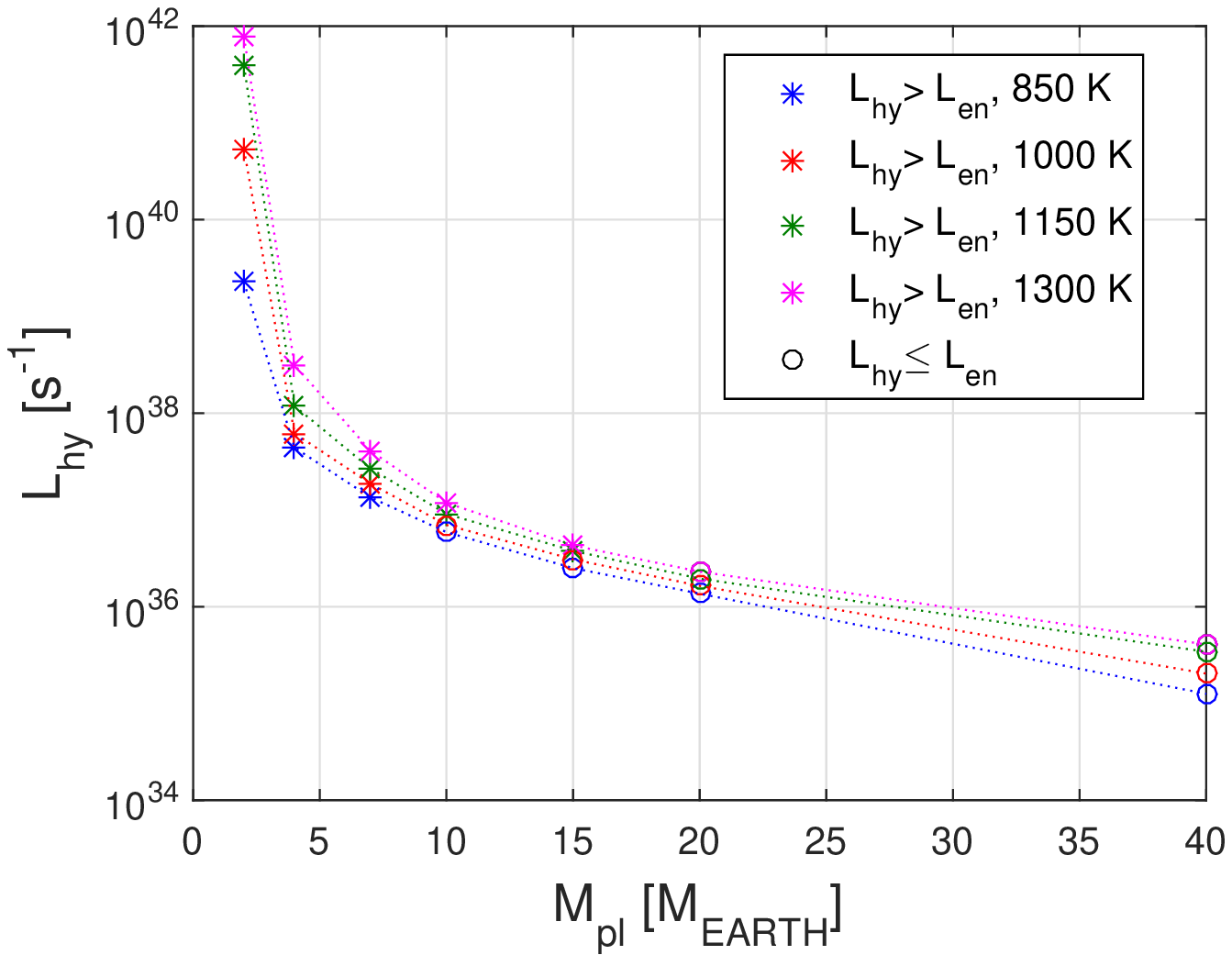}\\
  \includegraphics[width=\hsize]{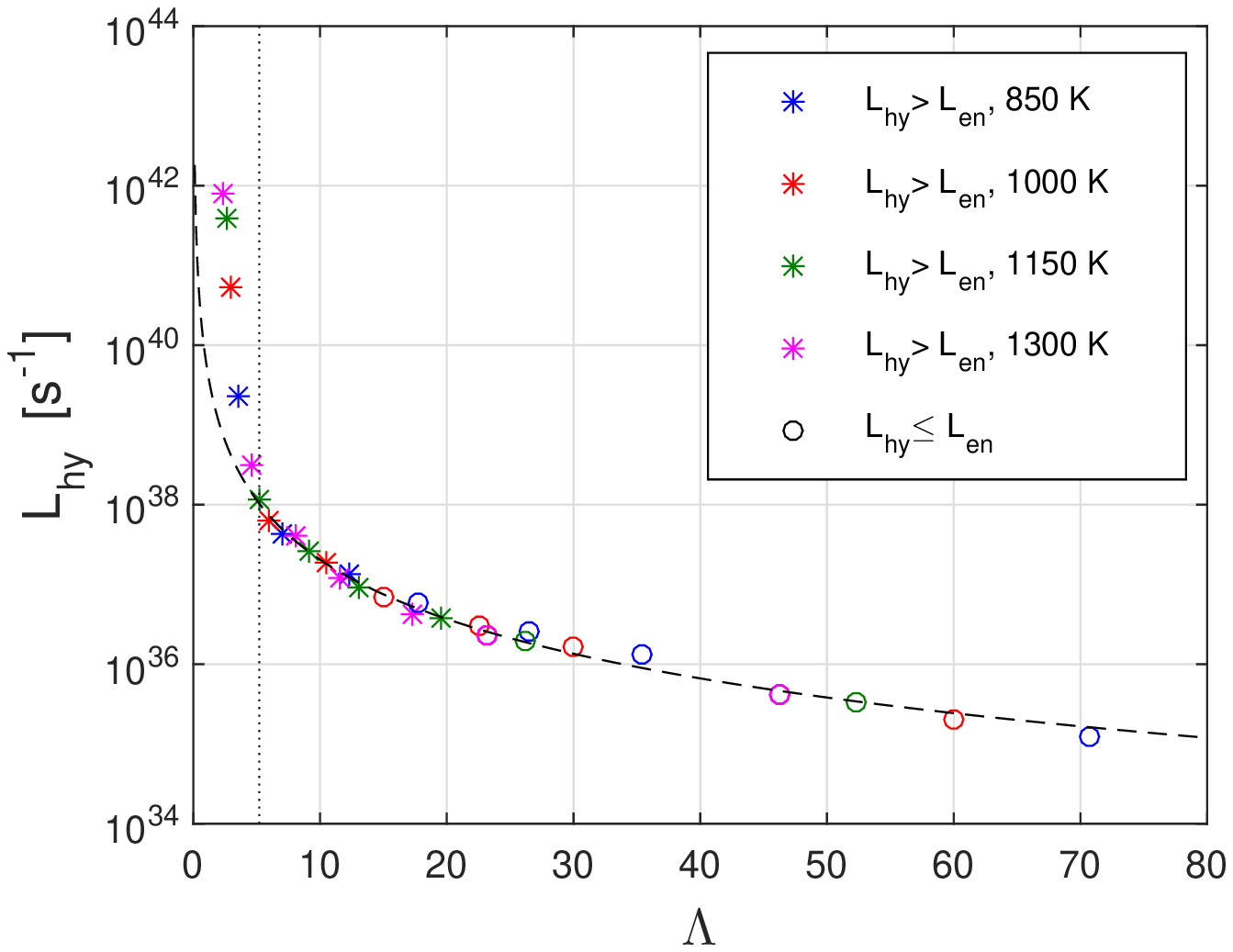}\\
  \caption{Hydrodynamic escape rates $L_{hy}$ as a function of planetary mass (top) and restricted Jeans escape parameter $\Lambda$ (bottom). Asterisks and circles mark models for which $\Lhy$ is higher and lower, respectively, than that estimated from the energy-limited formulation (see Section~\ref{sect:escape_rates_discussion}, for details). The dashed line in the bottom panel denotes Eq.~(\ref{eqn:approx}). The vertical dotted line indicates the minimum $\Lambda$ value for which Eq.~(\ref{eqn:approx}) is valid.}
  \label{fig:lhy}
\end{figure}
%
\section{Results}\label{sec:results}
\subsection{Escape rates}
For each cell in the model grid, we calculate the properties of
the upper atmosphere including temperature, number density, and
bulk flow velocity as a function of radial distance. From these
quantities, we estimate the hydrodynamic hydrogen escape rate
$\Lhy$ as the flow of particles through the upper boundary
\begin{equation}\label{eqn:Lhy}
    \Lhy = 4\pi \roche^2n_1 V_1 \, ,
\end{equation}
where $n_1$ and $V_1$ are the number density and velocity at the
Roche radius $\roche$. Figure~\ref{fig:lhy} shows the values of
$\Lhy$ obtained from the simulations as a function of planetary
mass and restricted Jeans escape parameter
\begin{equation}
\Lambda = \frac{GM_{\rm pl}m_{\rm H}}{k_{\rm B}T_{\rm eq}R_{\rm pl}}\,.
\end{equation}
This last quantity corresponds to the value of the Jeans escape
parameter calculated for the observed planetary radius and mass
for the planet's equilibrium temperature, and considering atomic
hydrogen \citep{Fossati17}. This parameter is particularly useful
because it gives a rough measure of the planetary gravitational
potential versus the intrinsic planetary thermal energy (i.e.
excluding XUV heating), and hence it allows us to estimate the
stability of the atmosphere against its own thermal energy and
planetary gravity. The physical properties of the modelled planets
and the derived escape rates are listed in
Table~$3$.

As expected, the escape rates increase with decreasing planetary
mass. At high mass (i.e. $M_{\rm pl}$\,$>$\,10\,$M_{\oplus}$),
the escape is driven by the heating caused by the absorption of
the stellar XUV flux and the atmosphere is in the blow-off regime.
Heating caused by absorption of the stellar X-ray
flux contributes just a few percent of the total escape rates. For
lower masses, where $\Lhy$ increases steeply, the escape is driven
mostly by the planetary high thermal energy and low gravity. For
these planets, the atmosphere lies in the boil-off regime (see
Section~\ref{sect:escape_rates_discussion}).

\subsection{Temperature and velocity profiles}
%
\begin{figure*}[]
\begin{center}
  \includegraphics[width=9.1 cm]{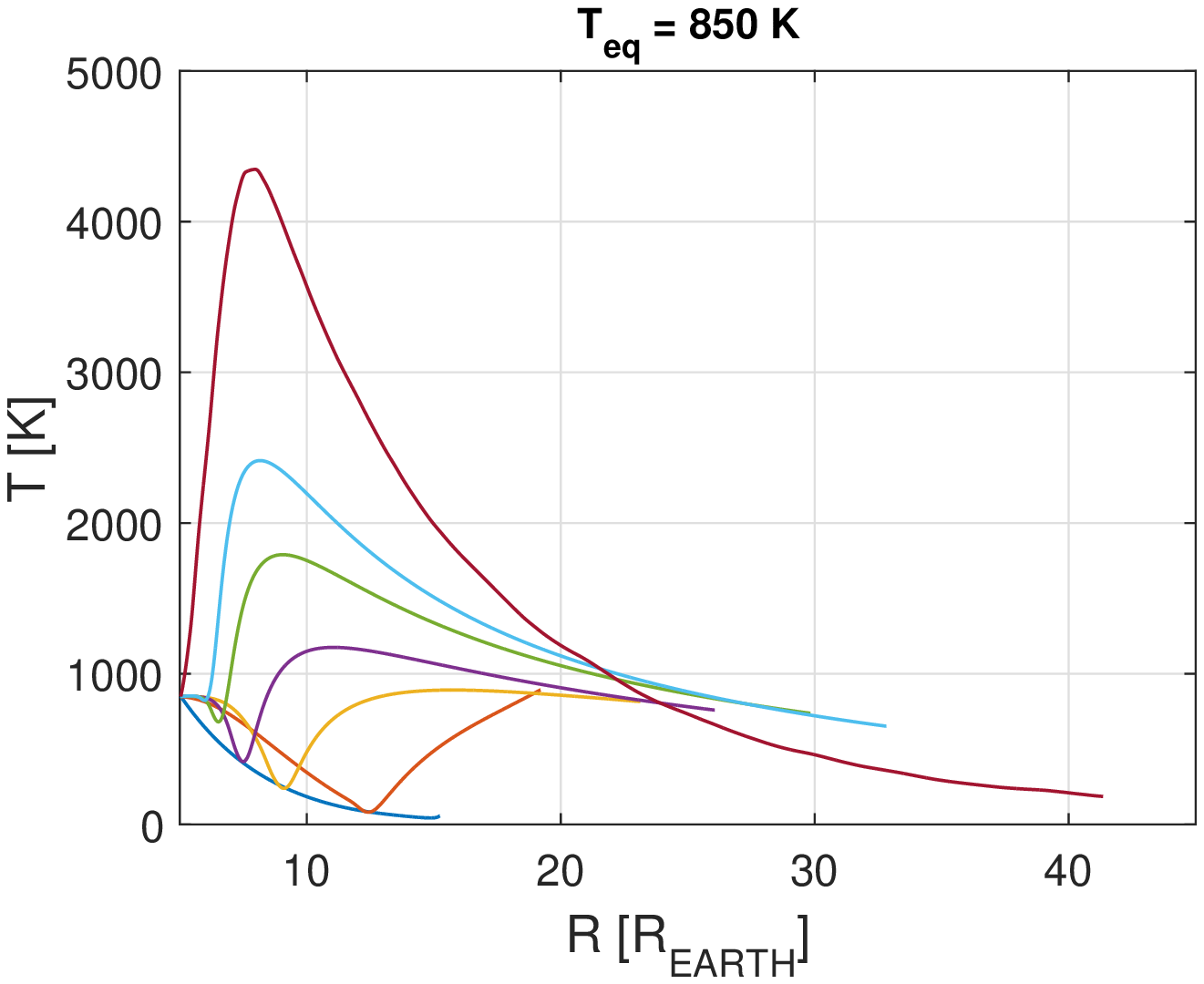}
  \includegraphics[width=9.1 cm]{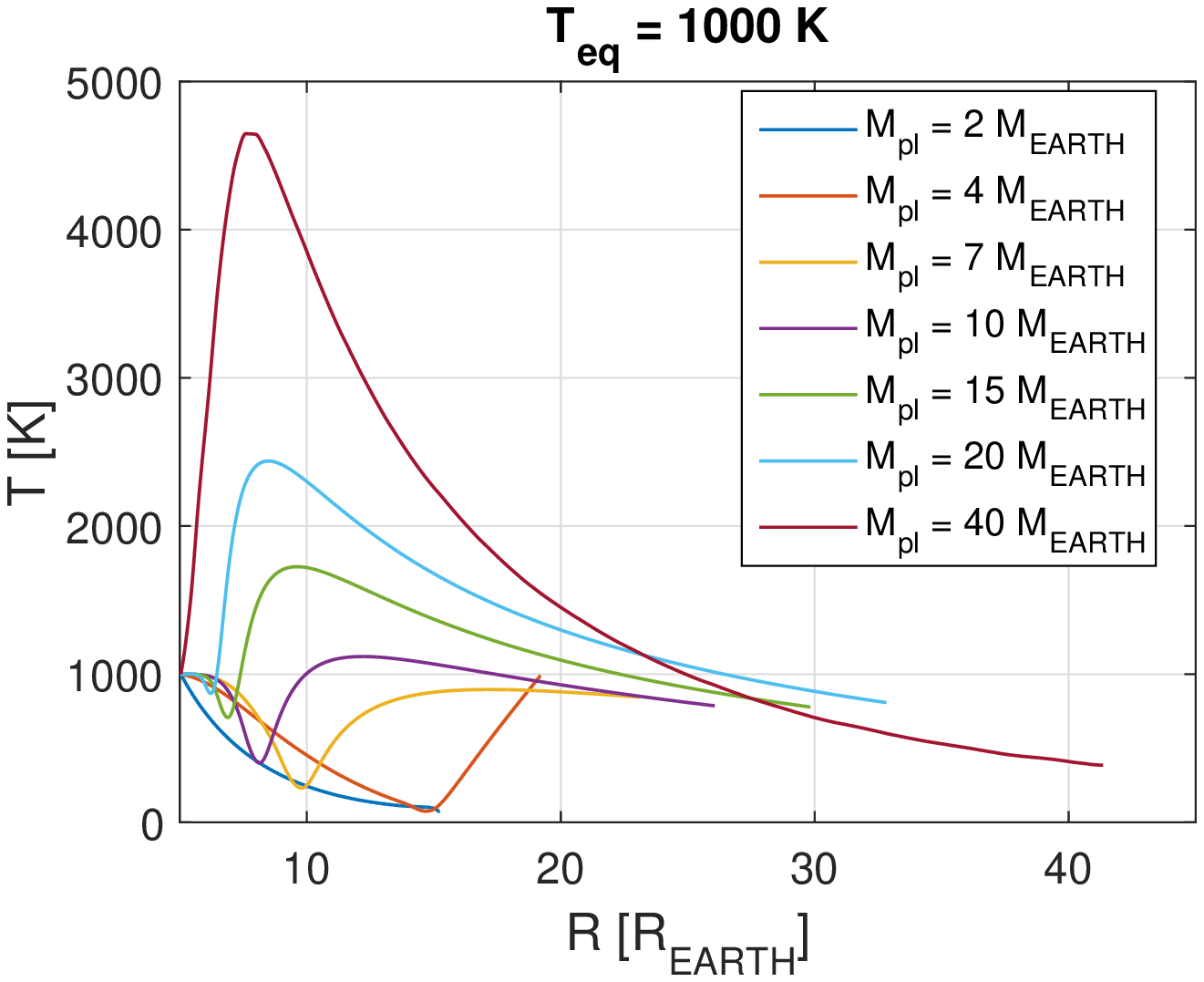}\\
  \includegraphics[width=9.1 cm]{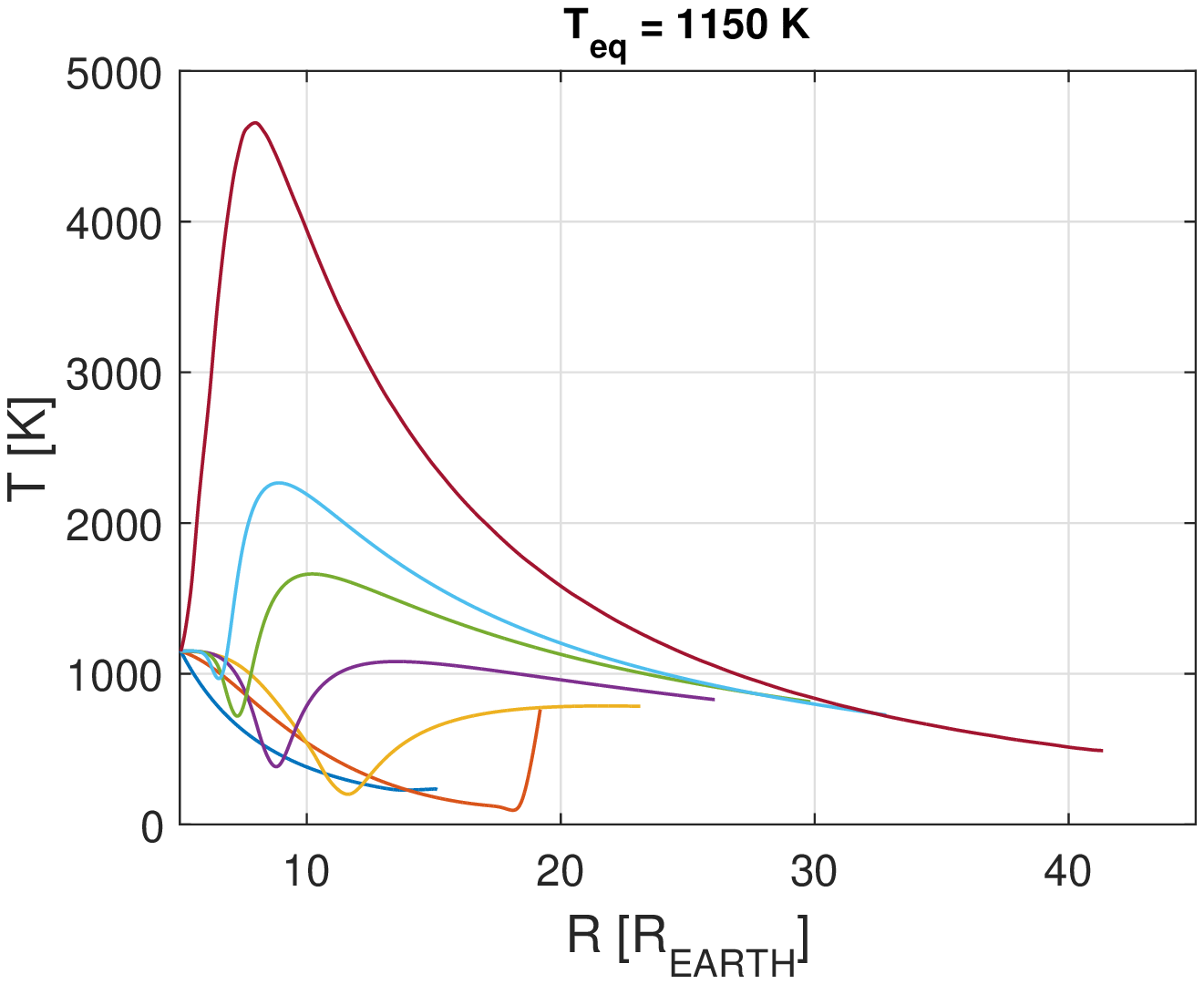}
  \includegraphics[width=9.1 cm]{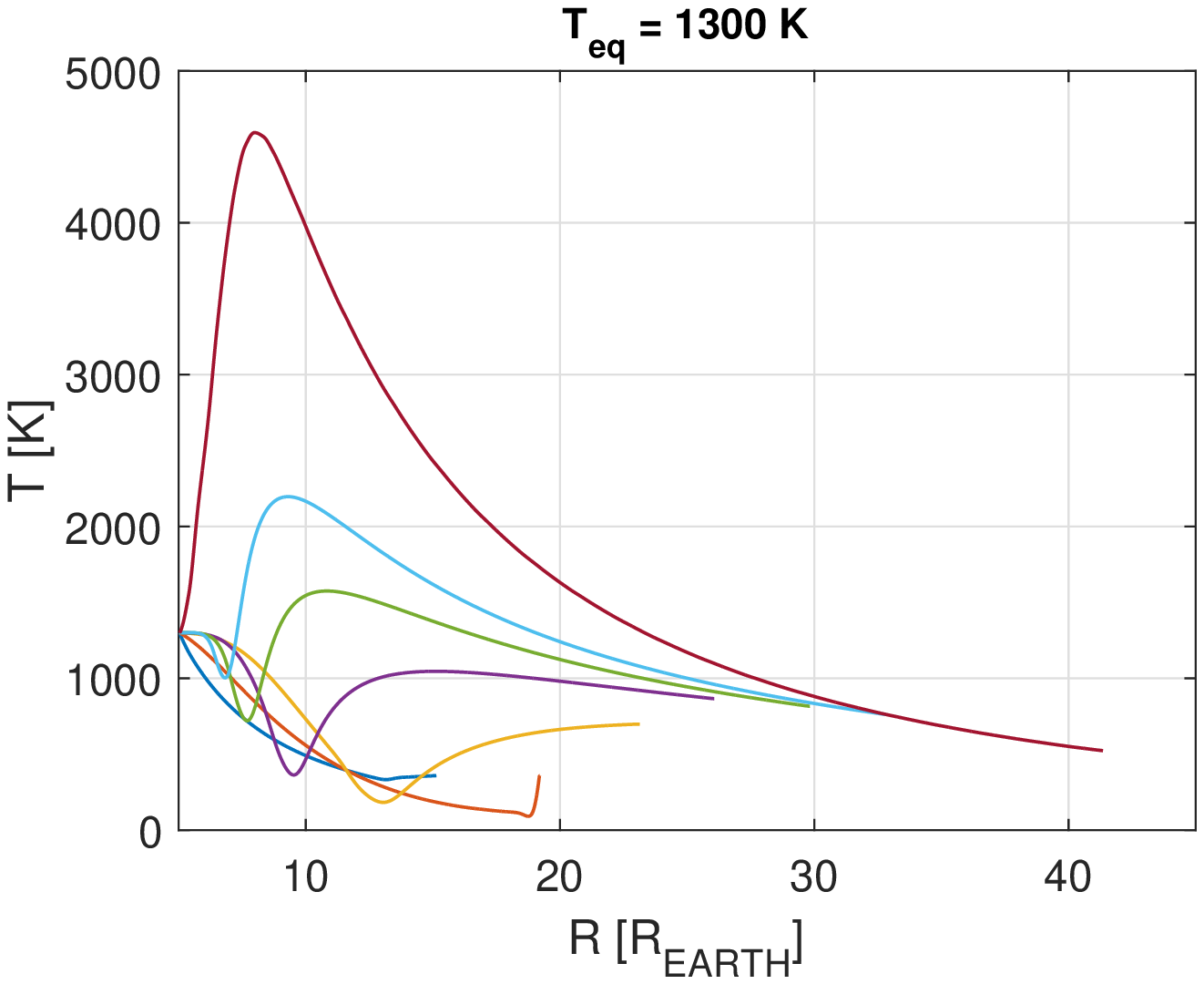}\\
  \caption{Temperature profiles of the modelled planets as a function of planetary mass and equilibrium temperature. For planets with a mass between 7 and 15\,$M_{\oplus}$, the temperature profile presents a minimum close to $R_{\rm pl}$.}
  \label{fig:t_all}
  \end{center}
\end{figure*}
The atmospheric temperature profiles are mostly dependent on the
planetary mass, while the adopted \Teq\ value has only a small
influence (Fig.~\ref{fig:t_all}). For $M_{\rm pl} > 10 M_{\oplus}$
the models present efficient heating, with pronounced maxima,
caused by the absorption of the stellar XUV radiation. For the lowest
mass planets, the temperature profile is dominated by adiabatic
cooling, driven by the strong hydrodynamic expansion of the
atmosphere in the boil-off regime. For planets with intermediate
masses (i.e. 7--15\,$M_{\oplus}$), the temperature profile
presents a minimum close to $R_{\rm pl}$, the position and
amplitude of which decrease with increasing mass. We
discuss the origin of this temperature minimum in
Section~\ref{sec:discussion}.

For atmospheres dominated by adiabatic cooling, the temperature
decreases at approximately the same rate throughout the
atmosphere, independently of \Teq. For the profiles
presenting XUV heating in the atmosphere, the value of the maximum
temperature increases slightly with decreasing temperature, with a
difference of about 250\,K between the coolest and hottest
temperature maxima.

{As stated in Section~\ref{sec:model}, the planetary atmosphere at
the lower boundary consists of pure molecular hydrogen. Due to the
XUV irradiation, hydrogen dissociates not far from the planetary
surface and the upper part of the atmosphere is thus dominated by
atomic hydrogen. The position of the dissociation barrier, the
thickness of the hydrostatic $\hh$ layers, and the atmospheric
temperature mainly depend on the planetary mass. The dissociation
appears to be more effective for the more massive planets, where
the stellar flux penetrates deeper into the planetary atmosphere.
Furthermore, the thickness of the pure molecular hydrogen layers
(i.e. where the other species contribute less than 1\%) varies
with planetary mass and equilibrium temperature. For $T_{eq} =
850$\,K, the thickness of these layers ranges between 1.3\,$R_{\rm
pl}$ for the lowest mass planets to 0.04\,$R_{pl}$ for the highest
mass planets. By increasing the equilibrium temperature from the
lowest to the highest value, the thickness of the molecular
hydrogen layer increases by about 30\%.}

\begin{figure*}[]
  \includegraphics[width=9.1 cm]{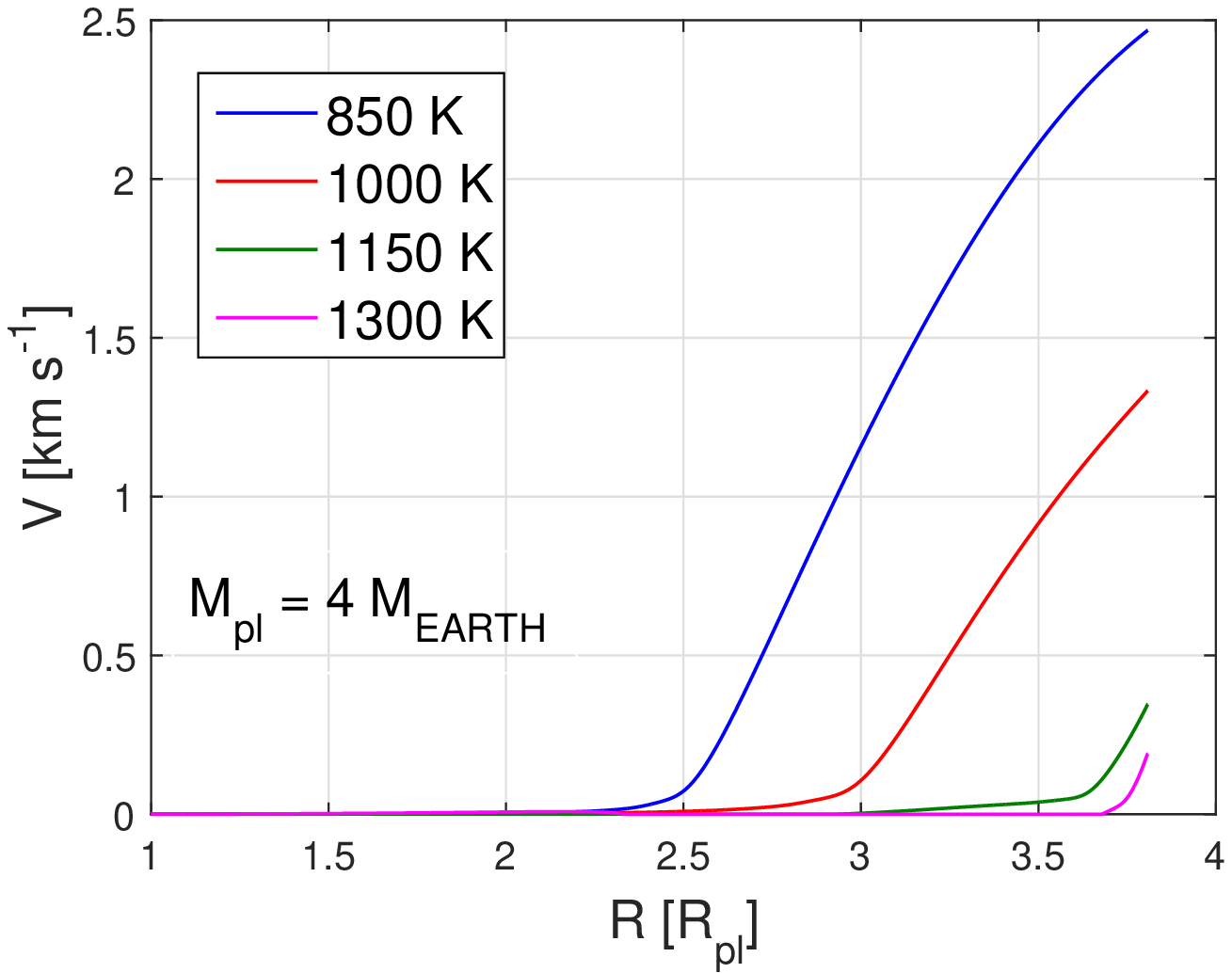}
  \includegraphics[width=9.1 cm]{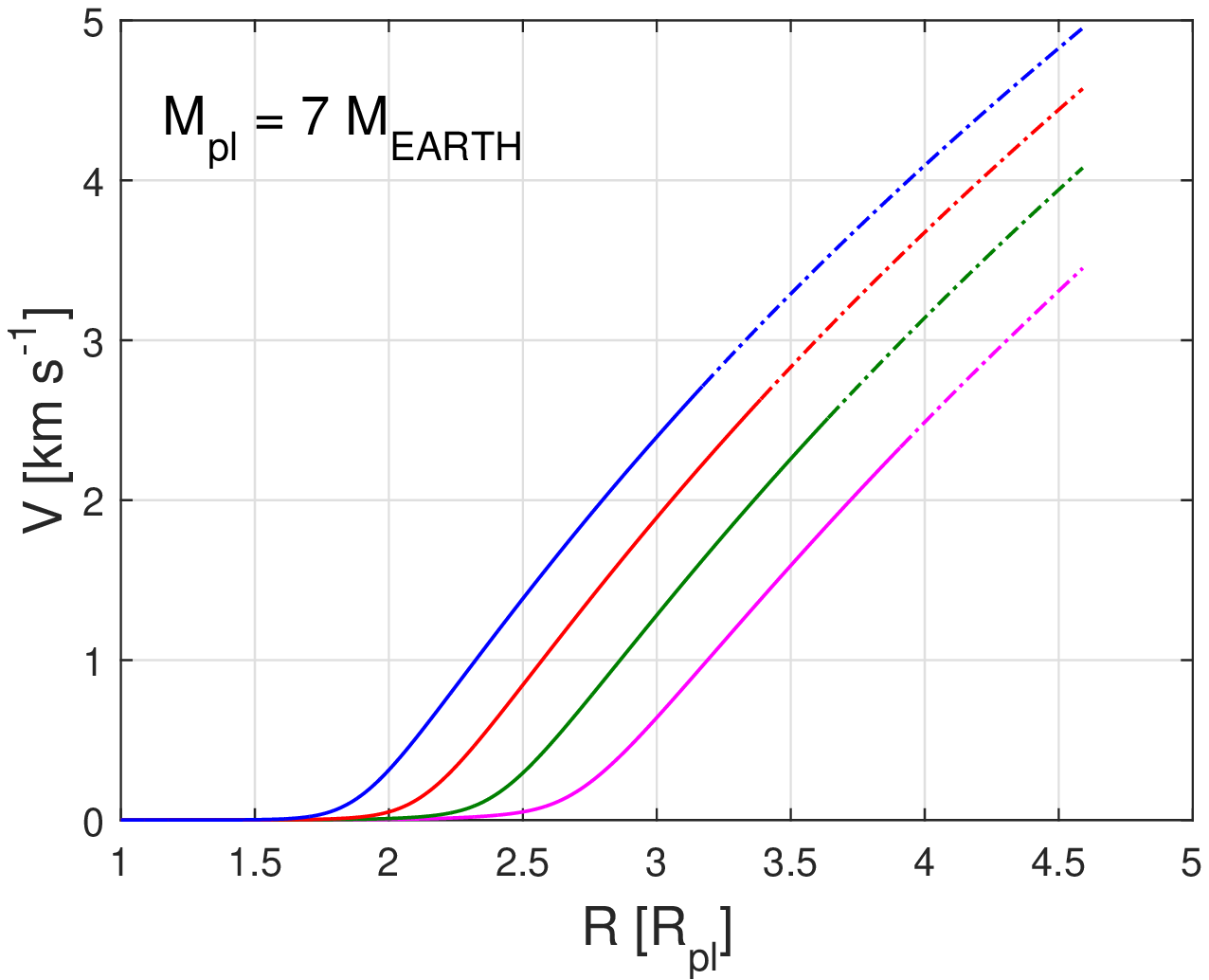}\\
  \includegraphics[width=9.1 cm]{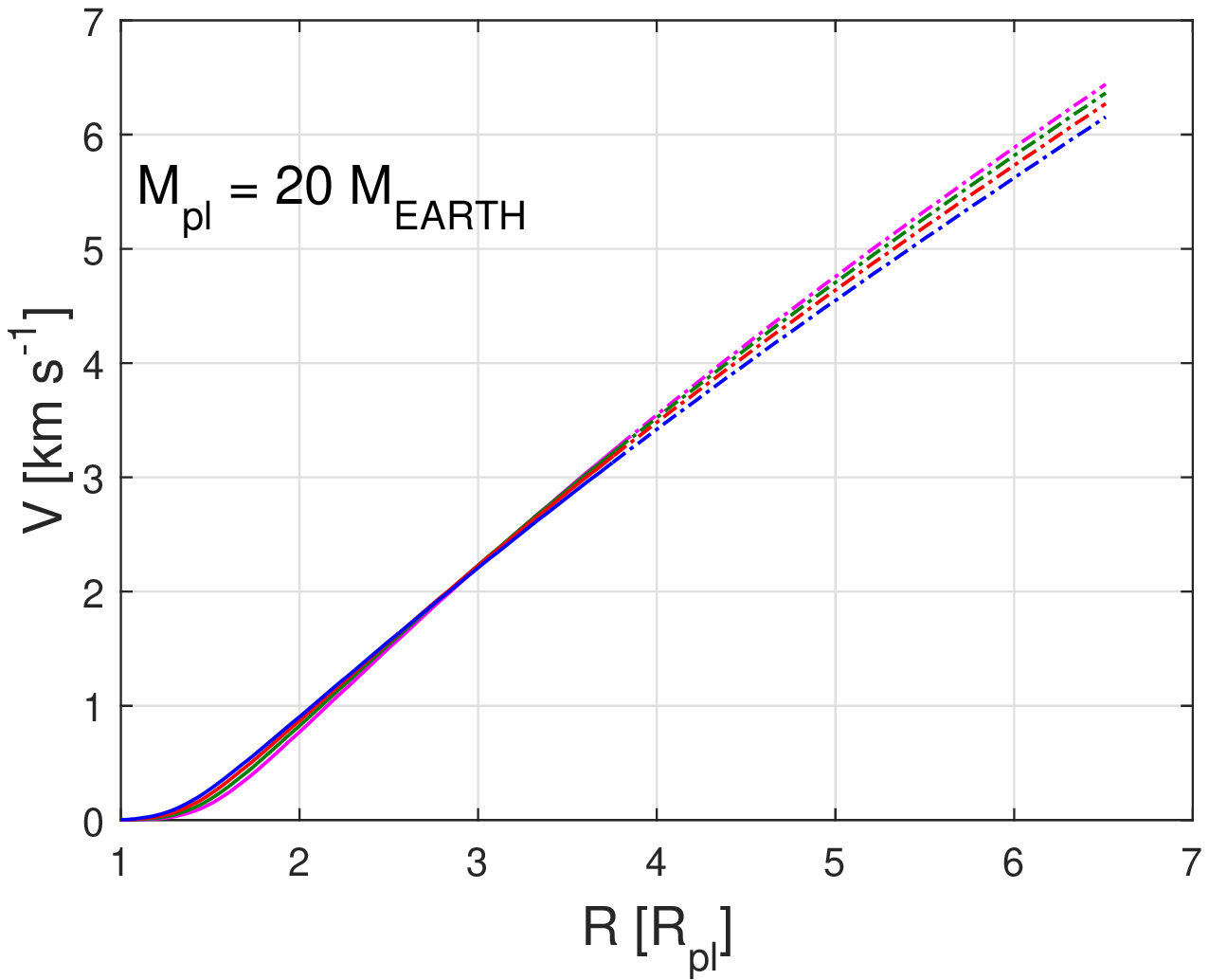}
  \includegraphics[width=9.1 cm]{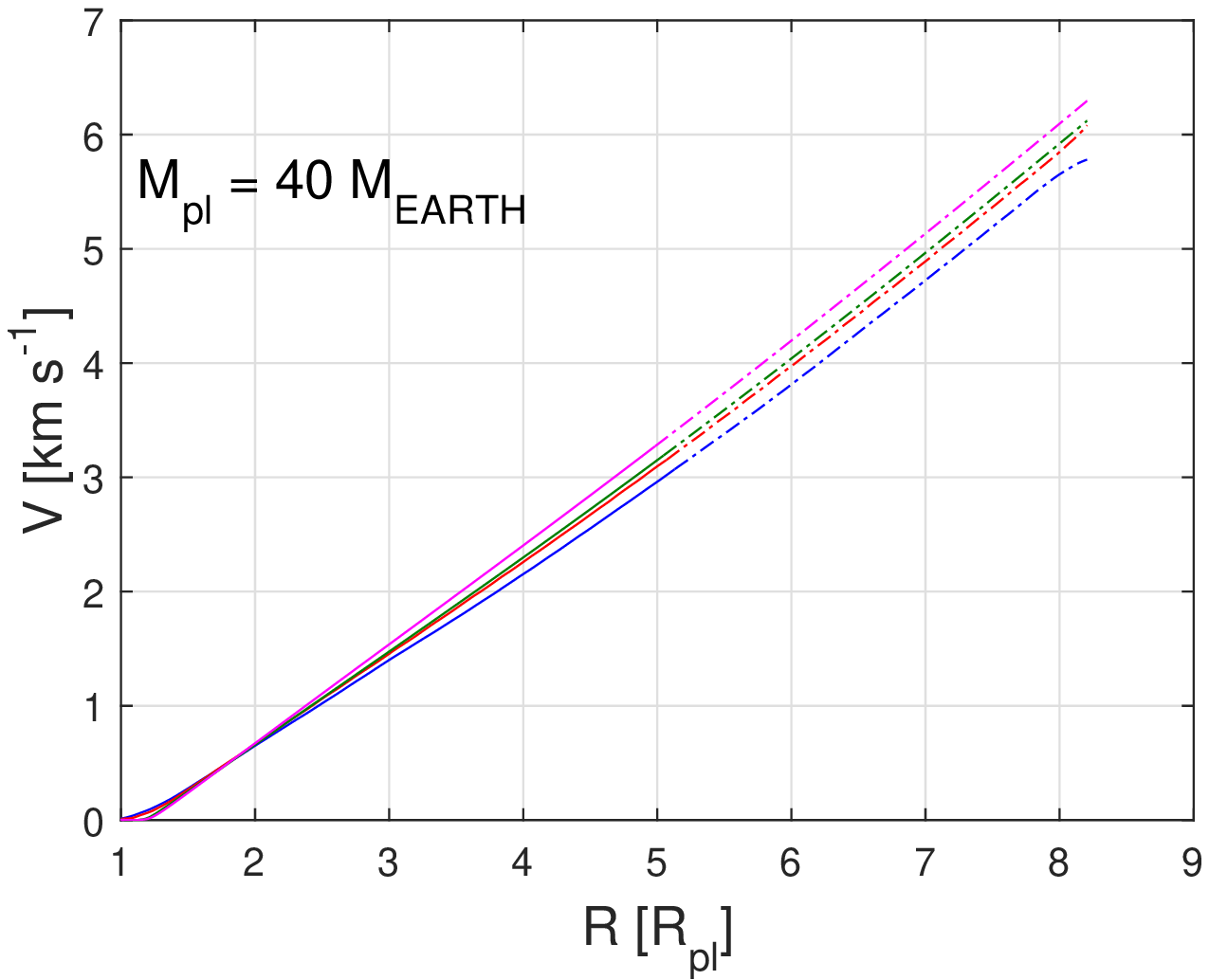}\\
  \caption{Gas velocity profiles for the planets with masses of 4 (top-left), 7 (top-right), 20 (bottom-left), and 40 (bottom-right)\,$M_{\oplus}$. The velocity profiles for planets with 10 and 15\,$M_{\oplus}$ are very similar to those of the 20\,$M_{\oplus}$ planet. The colours correspond to different values of the equilibrium temperature (blue: 850\,K; red: 1000\,K; yellow: 1150\,K; magenta: 1300\,K). The line style indicates whether a certain velocity is below (solid) or above (dash-dotted) the local sound speed. }
  \label{fig:V_all}
\end{figure*}
Figure~\ref{fig:V_all} shows the velocity profiles obtained for
the modelled planets. For the lowest mass planets, the bulk of the
atmosphere moves at a subsonic speed, with the local sound speed
being $V_{\rm T} = \sqrt{k_{\rm B}T/m_{\rm H}}$. For the other
planets, the upper part of the atmosphere becomes
supersonic above the position of the maximum
temperature. Similar to the behaviour observed for the temperature
profiles, the position at which the atmosphere becomes supersonic
gets closer to $R_{\rm pl}$ as the mass increases. The
sensitivity of both temperature and velocity profiles to
variations in \Teq\ decreases significantly with increasing mass:
\Teq\ only seems to play a role in shaping the atmospheric
profiles for the lower mass planets, for which the atmosphere is
in the boil-off regime.

\section{Discussion}\label{sec:discussion}
\subsection{Atmospheric escape regime}\label{sect:escape_rates_discussion}
%
\begin{figure}
    \includegraphics[width=\hsize]{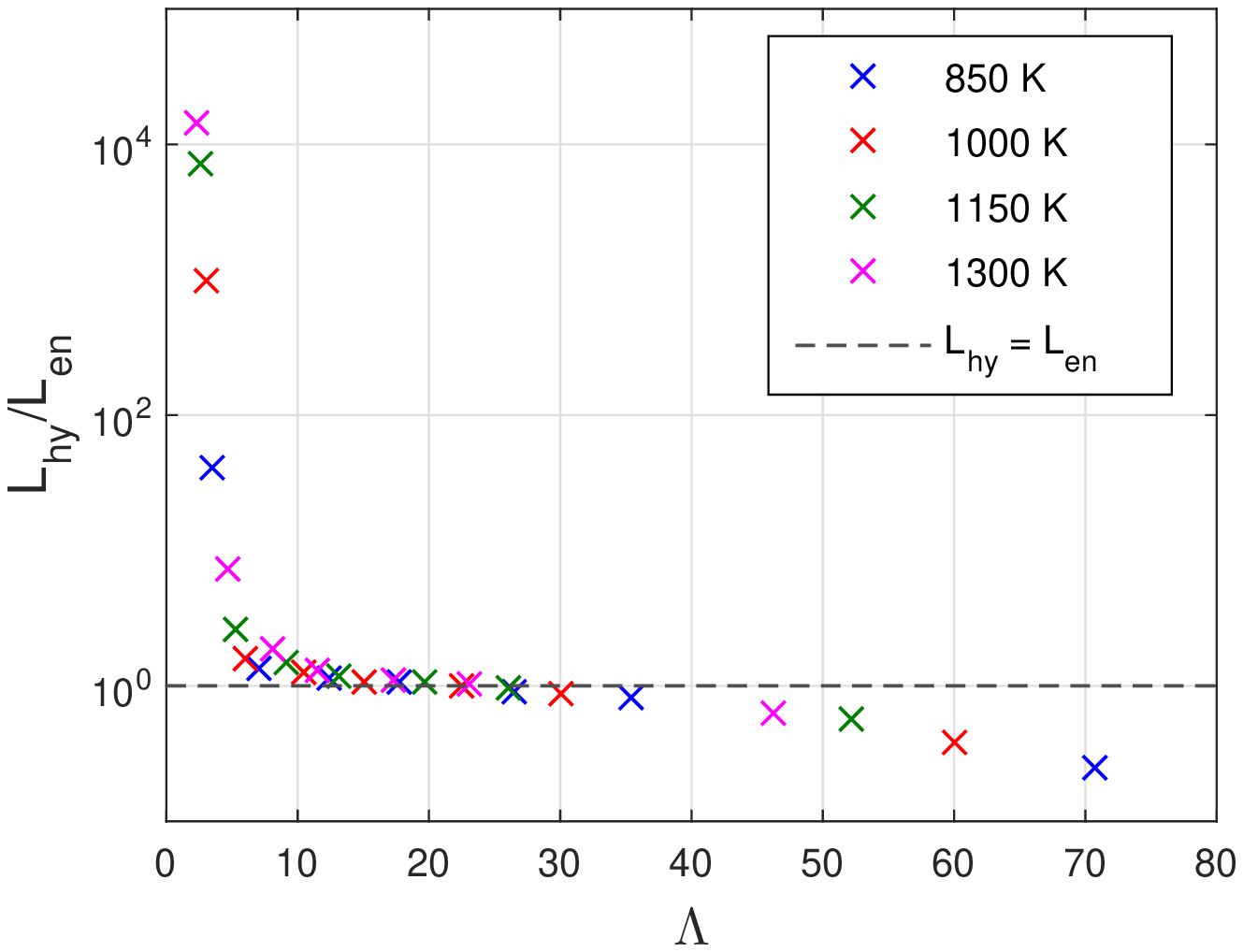}
    \caption{Ratio between the escape rates calculated with the hydrodynamic model $L_{\rm hy}$ and with the energy limited formulation $L_{\rm en}$ as a function of the restricted Jeans escape parameter $\Lambda$. The dashed horizontal line indicates the equality between $L_{\rm hy}$ and $L_{\rm en}$. The symbols are colour-coded as in the top panel and Fig.~\ref{fig:lhy}.}
    \label{fig:lhytole}
\end{figure}

We compare here the escape rates derived from the models with
those estimated through the energy-limited formulation
\citep{Watson1981,erkaev2007}

\begin{equation}\label{eqn:Len}
    \Len = \frac{R_{\rm pl}R_{\rm eff}^2\Fxuv}{GM_{\rm pl}K}\,,
\end{equation}
where $\Fxuv$ is the stellar XUV flux incident on the planet,
$\Reff$ is the radius at which most of the XUV flux is absorbed,
and $K$ is a term accounting for Roche lobe effects (Table~$3$).
The value of $\Reff$ is calculated as
\begin{equation}\label{eqn:Reff}
\Reff = R_{\rm pl}\sqrt{1+2\int_1^{\infty}{\left(1-\frac{J_{\rm XUV}(r,\frac{\pi}{2})}{\Fxuv}\right)r}dr}\,.
\end{equation}
%
In Eq.~(\ref{eqn:Reff}), $J_{\rm XUV}$ corresponds to the stellar
XUV flux inside the atmosphere, which depends on $r$ and on the
spherical angle \citep[see][for more
details]{erkaev2007,erkaev2015}. The $\Reff$ values correspond
roughly to the positions of the temperature maxima shown in
Fig.~\ref{fig:t_all} and hence $\Reff$ increases with decreasing
$\Lambda$. For the lowest mass planets, where the temperature
decreases adiabatically, $\Reff$ lies instead above the Roche
radius, which is the upper boundary of our calculations. For such
planets, we therefore consider $\Reff$ to be equal to $\roche$,
hence slightly underestimating the $\Len$ values.

Figure~\ref{fig:lhytole} shows the $\Lhy$/$\Len$ ratio as a
function of $\Lambda$. This plot indicates that for $\Lambda$
values ranging between 10 and 40, the two escape rates are roughly
similar, while for lower $\Lambda$ the escape rates calculated
using the hydrodynamic model are significantly higher than those
derived using Eq.~(\ref{eqn:Len}). This agrees with the findings of
\citet{Fossati17}, who carried out a similar analysis, but for 5
and 10\,$M_{\oplus}$ planets subject to much smaller amounts
(i.e. factor of 10 to 1000) of XUV irradiation. The atmospheres of
planets with the lowest $\Lambda$ values lie therefore in the
boil-off regime. This is also shown, for example, by the strong
dependence of the escape rates on \Teq\ for these
planets (see Fig.~\ref{fig:lhy}).

Given the monotonic increase of the escape rates with decreasing
$\Lambda$, we fit the $\Lhy$ values with a logarithmic function of
$\Lambda$, obtaining
\begin{equation}\label{eqn:approx}
    \ln{L_{\rm hy}} = 91.7-2.5\ln{\Lambda}\,,
\end{equation}
where the fit was done excluding the planets for which $\Reff$
lies above $\roche$. It is important to remark that this fit is
valid just for planets with $\Lambda$ greater than about five and for
the stellar parameters, including the XUV fluxes, adopted for
these calculations. Higher or lower XUV fluxes would lead to
higher or lower escape rates, respectively.

\subsection{Impact of the uncertainties of the stellar parameters}\label{sect:uncertainties_parameters}
As mentioned in Section~\ref{sec:param}, there is a significant
difference in the stellar mass estimated by \citet{david2016} and
\citet{mann2016}. This study is based on the stellar mass of
0.56\,$M_{\odot}$ derived by \citet{mann2016} and we discuss here
the impact of this choice on the results. We re-calculated the
grid of planetary upper atmosphere models considering a stellar
mass of 0.3\,$M_{\odot}$ \citep{david2016}. We found that, because
of the decreased gravitational influence of the star on the
planet, hence of the larger planetary Roche lobe, the escape rates
for the lower mass planets are about half of those obtained
considering the higher mass star. For the higher mass planets,
instead, the modification of the gravitational potential has no
effect on the escape rates.

The lower stellar mass leads also to lower stellar XUV fluxes.
{This is due to} the dependence of the X-ray luminosity on the
stellar mass, such that $L_{\rm X,0.56} \approx 1.8 L_{\rm
X,0.3}$, where the subscript indicates the stellar mass. {Thus,
when} considering the lower stellar mass, the escape rates
decrease by about 36\% compared to those derived considering the
higher stellar mass.

A further, closely related, uncertainty is that of the stellar
bolometric luminosity, which enters also in the calculation of the
XUV fluxes. By considering for example the estimate of $L_{\rm
bol}$ based on the stellar age \citep{Preibisch2005}, the XUV
fluxes would increase by a factor of two. This would in turn lead
to an increase of 40\% in the escape rates for the most massive
planets. These considerations indicate that the major
uncertainties in the stellar parameters lead to escape rates that
are within a factor of two of those presented in
Section~\ref{sec:results}.
\subsection{Possible impact of the modelling formalism}
{We discuss here how the results depend on the major modelling
assumptions. One dimensional models are widely used for exoplanet
studies, particularly with respect to upper atmosphere modelling
\citep[see
e.g.][]{mc2009,shaikhislamov2014,Salz2016,erkaev2016,erkaev2017,Fossati17,Lammer2016}.
An upgrade to 2D or 3D simulations is instead required to model
the interaction between the expanding planetary atmosphere with
the stellar wind, particularly when considering ion pick-up
processes related to planetary atmospheric escape. The mass loss
expected to occur because of charge exchange processes is,
however, about an order of magnitude smaller than the escape
caused by XUV heating \citep[see e.g.][]{kislyakova2014}, and
hence does not significantly affect our results. Furthermore, the
stellar wind does not play a significant role if the planetary
outflow is supersonic before the stellar-planetary interaction
occurs, because the stellar wind particles do not penetrate into
the deeper atmospheric layers. This is the case of K2-33b, except
possibly for the lowest considered masses. However, for these
cases, the planetary atmosphere fully escapes within 5\,Myr.}

{The presence of strong transversal planetary winds would
redistribute the thermal energy in the lower atmosphere from the
day to the night side, thus lowering the temperature at the lower
boundary of our simulations. Given the small dependence of our
results on the planetary equilibrium temperature, we expect that a
lower temperature by a few hundred degrees would not significantly
affect our results.}

{As was stated in Section~\ref{sec:physical_model}, we assume
solar metallicity when estimating the lower pressure boundary
$P_0$. We can therefore consider the effect of a different
metallicity varying the lower pressure boundary. We thus conducted
additional runs assuming a metallicity of 10$\times$ and
0.1$\times$ solar. We found that, at a temperature of 850\,K, the
escape rates change by about 5\% in the case of the $M_{\rm pl} =
20 M_{\oplus}$ planet and by about 15\% in the case of
7\,$M_{\oplus}$ planet. Other, more local, parameters, such as
temperature profiles, ionization rate, and position of the
dissociation barrier, vary within 1\%.}

{Finally, we tested the effect of the choice of chemical reaction
rates. We ran further tests on the two simulated planets mentioned
above (T$_{\rm  eq} = 850$\,K; M$_{\rm pl} = 7$ and $20
M_{\oplus}$) substituting the reaction rates given by
\citet{yelle2004} (used in this work) with those provided by the
UMIST Database for Astrochemistry (UDfA) ({\tt
http://udfa.ajmarkwick.net}). We found that the escape rates vary
by less than 0.5\%, with the largest variations (still below 5\%)
being for the ionization rate.}

\subsection{Evolutionary status}\label{sect:evolution}
We discuss here whether the derived escape rates, together with
the age of the system, can provide any constraint on the planet's
minimum mass. We do not aim to realistically trace
the evolution of the planet's atmosphere, which would require
more sophisticated models. We aim instead at providing an order of
magnitude estimation for the timescale expected for the complete
escape of a hydrogen-dominated atmosphere. By comparing it with
the system's age, we derive an estimate of the possible minimum
planetary mass. To this end, one needs to assume a value for the
current envelope mass fraction $f$, which is unknown. In the
literature, the maximum $f$ value that is commonly considered is
50\% \citep[e.g.][]{rogers2011,mordasini2012,owen2016b}. We
therefore consider possible envelope mass fractions of 50\%, 30\%,
10\%, and 1\%. Given also the weak dependence of the escape rates
on the equilibrium temperature, we consider the results
obtained for a \Teq\ value of 850\,K.

\begin{figure}
    \includegraphics[width=\hsize]{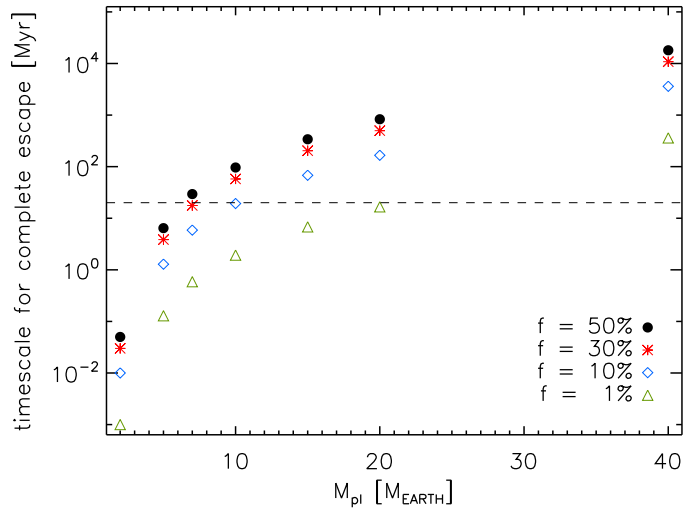}
    \caption{Timescale, in Myr, estimated for the complete escape of a hydrogen-dominated atmosphere as a function of planetary mass, assuming four different values of the atmospheric mass fraction $f$ of 50\% (black dot), 30\% (red asterisk), 10\% (blue diamond), and 1\% (green triangle). The horizontal dashed line indicates the maximum age (i.e. 20\,Myr) derived for the K2-33 system \citep{mann2016}.}
    \label{fig:evolution}
\end{figure}

Figure~\ref{fig:evolution} shows the timescale estimated for the
complete escape of a hydrogen-dominated atmosphere {(i.e. ratio
between atmospheric mass and mass-loss rate)} as a function of
planetary mass, for the four considered $f$ values. This plot
shows that, if the planet had a mass smaller than 7\,$M_{\oplus}$,
the planetary hydrogen-dominated envelope would have completely
escaped by now. However, such a low planetary mass would imply a
low bulk density, and therefore the presence of a thick
hydrogen-dominated atmosphere, which is in contradiction with the
previous statement. The result is that the planet cannot be less
massive than 7\,$M_{\oplus}$, regardless of the considered
envelope mass fraction. In the same way, if the planet has an $f$
value smaller than 30\% (or 10\%), one could also safely exclude
planetary masses smaller than 10\,$M_{\oplus}$ (or
15\,$M_{\oplus}$). For masses larger than 15--20\,$M_{\oplus}$,
the planet would keep its hydrogen-dominated atmosphere, even
considering small envelope mass fractions, hence ending up as a
hot Neptune.

These estimates do not consider that the stellar XUV flux, hence
the  atmospheric escape rates, will decrease with time. This will
lead to a slight decrease of the possible minimum mass. To
conclude, we can safely exclude the possibility that K2-33b is less massive than
7\,$M_{\oplus}$, or 10\,$M_{\oplus}$ if the envelope mass fraction
is below 30\%.

\subsection{Temperature minima}
%
\begin{figure}
  \includegraphics[width=\hsize]{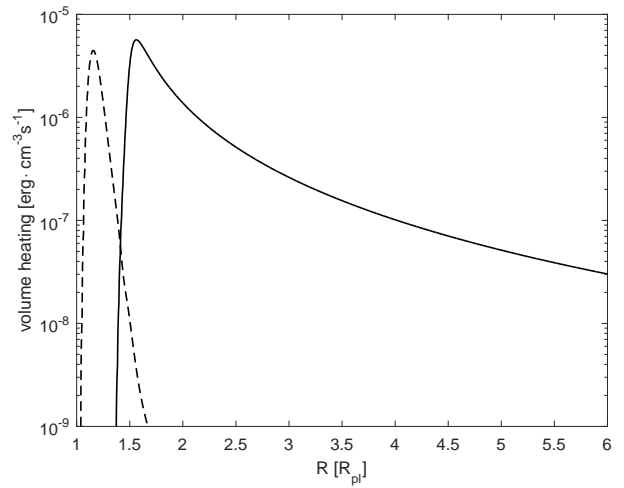}
  \caption{Volume heating rates of EUV (solid) and X-ray (dashed) stellar radiation for the 10\,$M_{\oplus}$ planet with an equilibrium temperature of 850\,K.}\label{fig:QQx}
\end{figure}
As mentioned in Section~\ref{sec:results}, the structure of the
temperature profiles shown in Fig.~\ref{fig:t_all} for planets
with masses between 7 and 15\,$M_{\oplus}$ is particularly
unusual. To understand the origin of the temperature minima, we
look at the effects of $\hhh$ cooling, X-ray heating, and EUV
heating.
\subsubsection{$\hhh$ cooling}
Interestingly, most of the cooling caused by $\hhh$ molecules is
located close to the position at which the temperature minimum
occurs. We therefore ran simulations for the 10\,$M_{\oplus}$
planet without considering $\hhh$ cooling, obtaining results that
were not appreciably different from those of the original runs
with the inclusion of $\hhh-$cooling. This result is in agreement
with those of \citet{shaikhislamov2014} and \citet{chadney2015}
who did similar tests for close-in Jupiter-mass planets.
\subsubsection{X-ray heating}
The X-ray heating rate is comparable to that produced by EUV, but,
because of the much smaller absorption cross-section ($\sigma_{\rm
X} \sim 0.5\times10^{-3}\sigma_{\rm EUV}$), the X-rays penetrate
significantly deeper into the atmosphere and deposit their energy in
a thin layer where the atmosphere is dominated by $\hh$ molecules
(Fig.~\ref{fig:QQx}). Already at 1.5\,$R_{\rm pl}$, X-ray heating
becomes negligible compared to that produced by the EUV.
Additional models run without the inclusion of the stellar X-ray
fluxes show that X-ray heating is not responsible for the
temperature minima either.

\subsubsection{EUV heating}
We further explored the effects of the intense EUV flux on the
temperature profiles by carrying out additional simulations with
the EUV fluxes reduced by a factor of 100 and 1000, compared to
what we estimated for K2-33. We found that the amplitude of the
minimum decreases in the first case and vanishes completely in the
second.

\begin{figure}
  \includegraphics[width=\hsize]{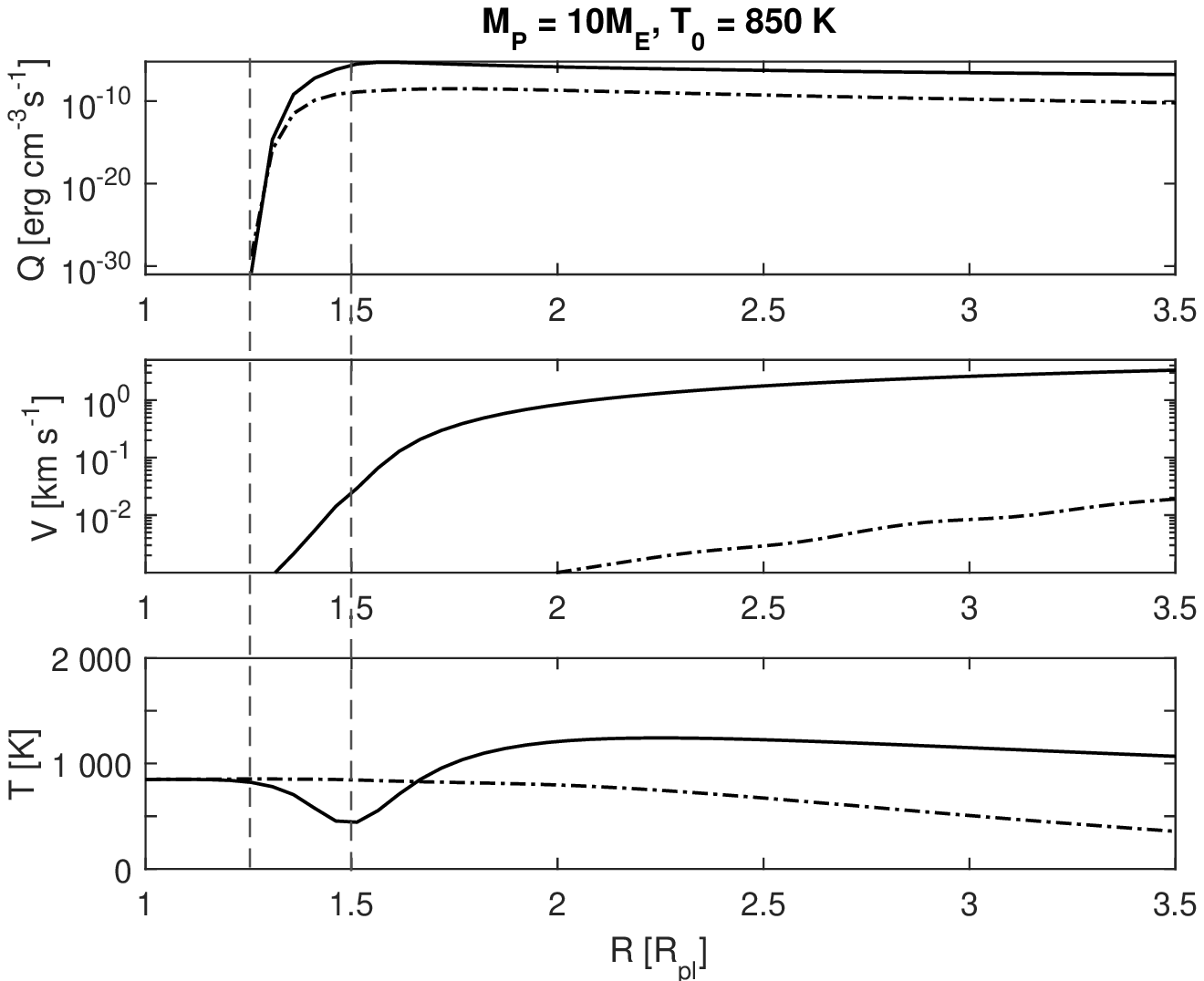}\\
  \includegraphics[width=\hsize]{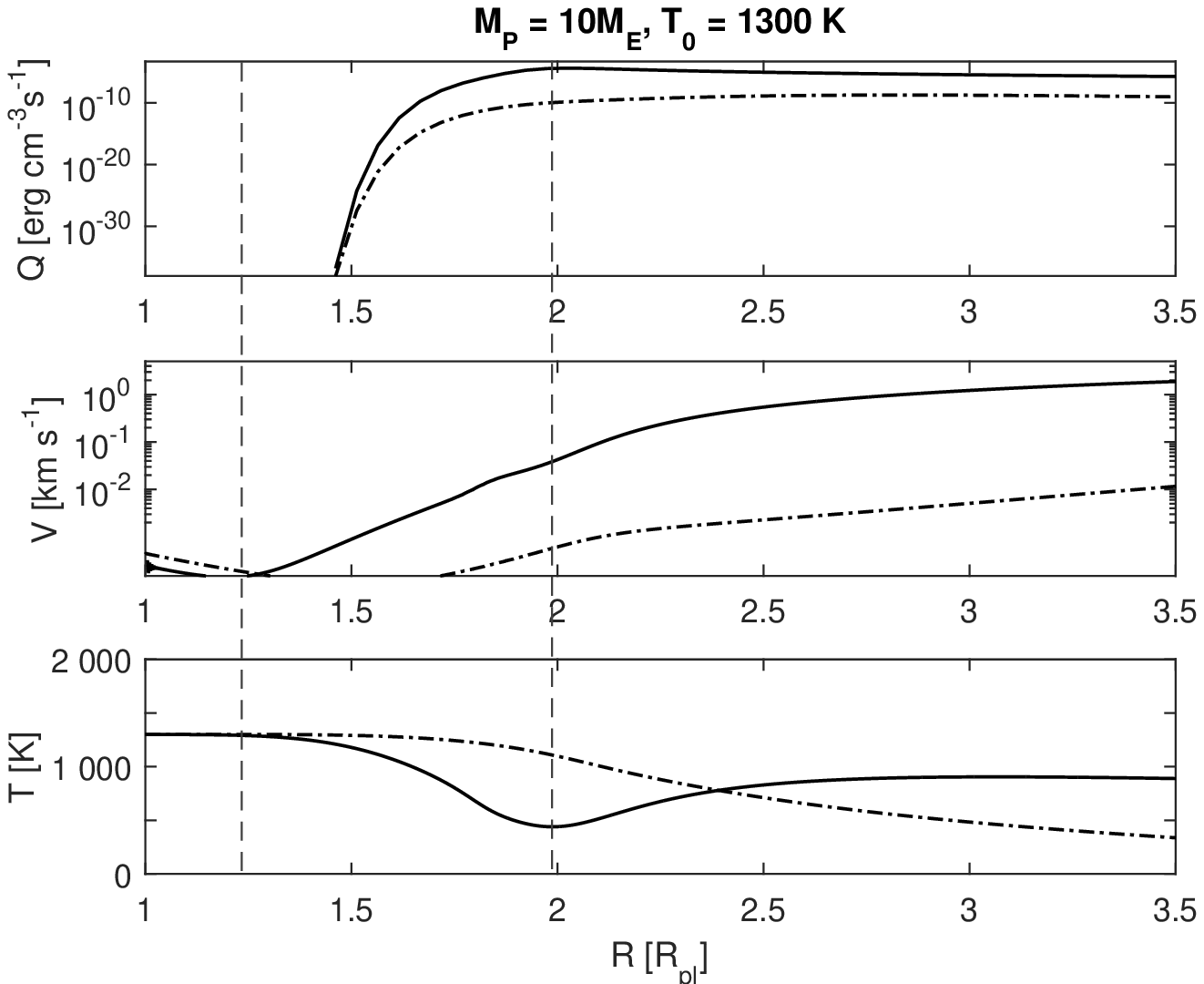}\\
  \caption{Top three panels: EUV volume heating rate (top), bulk velocity (middle), and temperature (bottom) profiles for the 10\,$M_{\oplus}$ planet with an equilibrium temperature of 850\,K. The solid line is for the planet irradiated by the XUV flux estimated for K2-33 and the dash-dotted line is for the same planet, but with a 1000 times weaker level of XUV irradiation. The vertical dashed lines mark the position at which the velocity begins to rise (left) and that of the temperature minimum for the most irradiated planet (right). Bottom three panels: same as the top three panels, but for a planet with an equilibrium temperature of 1300\,K. For the lower and higher temperature planets, $\Reff$ lies at about 2.3 and 3.0\,$R_{\rm pl}$, respectively.}
\label{fig:QVT_850}
\end{figure}

Figure~\ref{fig:QVT_850} shows the atmospheric profiles for
temperature, velocity, and EUV volume heating rate for the planets
with a mass of 10\,$M_{\rm \bigoplus}$ and an equilibrium
temperature of 850 and 1300\,K. For each planet, we {ran} two
simulations: one considering the EUV flux estimated for K2-33, and
one considering the same EUV flux reduced by a factor of 10$^3$.
In all cases, the atmospheres are characterized by adiabatic
cooling close to R$_{\rm pl}$, but only the planets subject to the
stronger EUV irradiation present a temperature minimum close to
the position of the peak of the volume heating rate.

The observed temperature minima are therefore caused by the
 extreme EUV heating. For the more irradiated planets, the EUV
heating rate is about five orders of magnitude higher than that of
the less irradiated planets, which also leads to the formation of
a hydrodynamic flow much closer to $R_{\rm pl}$. This results in a
strong adiabatic cooling caused by the fast gas flow, which leads
to a local decrease in the temperature. However, when the heating
reaches values of about
1.5--3$\times$10$^{-6}$\,erg\,cm$^{-3}$\,s$^{-1}$, it dominates
the adiabatic cooling, producing the observed temperature minimum.

\section{Conclusions}\label{sect:conclusions}
The young K2-33b planet offers the unique opportunity to study the
very early phases of planetary evolution at a time when the
planetary atmosphere is subject to extremely intense high-energy
stellar fluxes. This is therefore the time when atmospheric escape
is strongest and shapes the planetary envelope, setting it onto
its final evolutionary path. We employed an upper atmosphere
hydrodynamic model to study the physical characteristics of the
planetary envelope and infer the escape rates. In order to more
adequately model such a young system, we included in the model
X-ray heating and $\hhh$ cooling, in addition to the processes
included in the model in previous studies (\citealt{erkaev2016}). In the absence of a measured
planetary mass, we consider masses ranging between 2 and
40\,$M_{\oplus}$, while the planetary radius was kept fixed to the
measured value.

We found that for values of the restricted Jean escape parameter
$\Lambda$ smaller than 10, the escape rates are considerably
larger than those derived from the energy-limited formula. For
these planets, the escape is driven by the high intrinsic thermal
energy of the atmosphere and low planetary gravity, while XUV
heating plays a negligible role. For planets with a $\Lambda$
value between 10 and 40, the hydrodynamic escape rates are
comparable to those provided by the energy limited formula, while
for larger $\Lambda$ values they are lower.

For the lowest mass planets, the temperature profile is
characterized by adiabatic cooling, caused by the strong
hydrodynamic outflow of the atmosphere. For the highest mass
planets, the temperature profiles present a maximum close to the
position of maximum absorption of the stellar EUV flux (we find
that X-ray heating plays a negligible role compared to EUV
heating). For the intermediate mass planets, we find instead the
presence of a local minimum close to the planetary radius.
Additional dedicated simulations showed that this feature is
caused by the combination of adiabatic cooling and the absorption
of the strong stellar EUV fluxes.

Finally, we roughly estimate whether the derived escape rates,
together with the age of the system, can provide any constraint on
the minimum mass possible for the planet. We found that for masses smaller than
7\,$M_{\oplus}$, the planet would have completely lost its
hydrogen-dominated envelope, which is in contradiction with the
low average density that would result from a low planetary mass.
We therefore place a lower limit of 7\,$M_{\oplus}$ on the
planetary mass.

\begin{table*}[t]\label{tab:tab2}
\centering \caption{Reactions and relative cross-sections employed
in the model.}
\begin{tabular}{|M{4 cm}|M{6 cm}|M {4 cm}|N}
  \hline
  $H \rightarrow H^+ + e$  & $\nu_{H} =  5.9\times 10^{-8}\phi_{EUV} s^{-1}$ &
  \citet{storey1995}&  \\[18 pt]
  \hline
  $\hh \rightarrow \hhp + e$ & $\nu_{\hh} = 3.3\times 10^{-8}\phi_{EUV} s^{-1}$ & \citet{mc2009} &\\[18 pt]
  \hline
  $H^{+} + e \rightarrow H$ & $\alpha_{H} = 4\times 10^{-12}(300/T)^{0.64} cm^{3}s{-1}$ & \citet{yelle2004}&\\[18 pt]
  \hline
  $\hhp + e \rightarrow H + H$ & $\alpha_{\hh} = 2.3\times 10^{-8} (300/T)^{0.4} cm^3 s^{-1}$ & \citet{yelle2004} &\\[18 pt]
  \hline
  $\hh \rightarrow H + H$ & $\nu_{diss} = 1.5\times 10^{-9} e^{(-49000/T)}$ & \citet{yelle2004} &\\[18 pt]
  \hline
  $H + H \rightarrow \hh$ & $\gamma_H = 8.0\times 10^{-33} (300/T)^{0.6}$ & \citet{yelle2004} &\\[18 pt]
  \hline
  $H + e \rightarrow H^+$ & $\nu_{Hcol} = 5.9\times 10^{-11}T^{1/2}e^{(-157809/T)}$ & \citet{black1981} &\\[18 pt]
  \hline
  $\hhp + \hh \rightarrow \hhh + H$ & $\gamma_{\hh} = 2\times 10^{-9}$ & \citet{yelle2004} &\\[18 pt]
  \hline
  $\hhh + H \rightarrow \hhp + \hh$ & $\gamma_{\hh} = 2\times 10^{-9}$ & \citet{yelle2004} &\\[18 pt]
  \hline
  $\hhh + e \rightarrow \hh + H$ & $\alpha_{\hhh1} = 2.9\times 10^{-8}(\frac{300}{T_e})^{0.65}$ & \citet{yelle2004} &\\[18 pt]
  \hline
  $\hhh + e \rightarrow H + H + H$ & $\alpha_{\hhh2} = 8.6\times 10^{-8}(\frac{300}{T_e})^{0.65}$ & \citet{yelle2004} &\\[18 pt]
  \hline
\end{tabular}
\end{table*}

\begin{table*}[t]
\label{tab:tab3}
\centering
\caption{Physical properties of the modelled planets and escape and mass-loss rates derived from the simulations. The mass-loss rates are given in $g\,s^{-1}$ and in $M_{\oplus}\,Gyr^{-1}$.}
\begin{tabular}{c|c|c|c|c|c|c|c|c|c|c}
  \hline
  $M_{\rm pl}$ & $T_{\rm eq}$ &  $\Lambda$ & $L_{\rm hy}$ & $L_{\rm en}$ &
  $\dot{M}_{\rm hy}$ & $\dot{M}_{\rm en}$ &  $\dot{M}_{\rm hy}$ & $\dot{M}_{\rm en}$ & $R_{\rm roche}$ & $R_{\rm eff}$ \\
  $[M_{\oplus}]$ & [K] &  & $[10^{35}s^{-1}]$
  & $[10^{35}s^{-1}]$ &~$[10^{11}gs^{-1}]$ &~$[10^{11}gs^{-1}]$
  &~$[M_{\oplus}Gyr^{-1}]$&~$[M_{\oplus}Gyr^{-1}]$
  &~$[R_{\rm pl}]$ &~$[R_{\rm pl}]$ \\
  \hline
$2.0$   & $850$  & $3.5$  & $22654.8$  & $585.1$ & $37892.4$   & $978.6$ & $20008.9$   & $516.7$   & $3.02$ & $3.02$ \\
$4.0$   & $850$  & $7.1$  & $440.4$    & $346.8$ & $736.6$     & $580.0$ & $389.0$     & $306.3$   & $3.81$ & $3.29$ \\
$7.0$   & $850$  & $12.4$ & $134.7$    & $126.1$ & $225.3$     & $210.8$ & $119.0$     & $111.3$   & $4.59$ & $2.62$ \\
$10.0$  & $850$  & $17.7$ & $58.7$     & $59.6$  & $98.3$      & $99.6$  & $51.9$      & $52.6$    & $5.17$ & $2.16$ \\
$15.0$  & $850$  & $26.5$ & $25.0$     & $29.3$  & $41.8$      & $49.0$  & $22.1$      & $25.9$    & $5.92$ & $1.85$ \\
$20.0$  & $850$  & $35.3$ & $13.6$     & $17.8$  & $22.8$      & $29.8$  & $12.0$      & $15.7$    & $6.51$ & $1.67$ \\
$40.0$  & $850$  & $70.7$ & $1.3$      & $4.7$   & $2.1$       & $7.9$   & $1.1$       & $4.2$     & $8.20$ & $1.21$ \\
$2.0$   & $1000$ & $3.0$  & $536878.1$ & $585.3$ & $897982.4$  & $979.0$ & $474176.6$  & $517.0$   & $3.02$ & $3.02$ \\
$4.0$   & $1000$ & $6.0$  & $618.6$    & $418.4$ & $1034.6$    & $699.8$ & $546.3$     & $370.0$   & $3.81$ & $3.61$ \\
$7.0$   & $1000$ & $10.5$ & $186.7$    & $156.8$ & $312.3$     & $262.3$ & $164.9$     & $138.5$   & $4.59$ & $2.93$ \\
$10.0$  & $1000$ & $15.0$ & $69.8$     & $70.4$  & $116.7$     & $117.8$ & $61.6$      & $62.2$    & $5.17$ & $2.35$ \\
$15.0$  & $1000$ & $22.5$ & $30.8$     & $33.1$  & $51.5$      & $55.4$  & $27.2$      & $29.2$    & $5.92$ & $1.97$ \\
$20.0$  & $1000$ & $30.0$ & $16.4$     & $19.8$  & $27.4$      & $33.0$  & $14.5$      & $17.4$    & $6.51$ & $1.76$ \\
$40.0$  & $1000$ & $60.1$ & $2.0$      & $5.7$   & $3.4$       & $9.6$   & $1.8$       & $5.0$     & $8.20$ & $1.34$ \\
$2.0$   & $1150$ & $2.6$  & $3950771.2$& $585.3$ & $6608059.9$ & $979.0$ & $3489363.6$ & $517.0$   & $3.02$ & $3.02$ \\
$4.0$   & $1150$ & $5.2$  & $1184.9$   & $462.4$ & $1981.8$    & $773.5$ & $1046.5$    & $408.4$   & $3.81$ & $3.80$ \\
$7.0$   & $1150$ & $9.1$  & $267.2$    & $192.7$ & $446.9$     & $322.4$ & $236.0$     & $170.2$   & $4.59$ & $3.24$ \\
$10.0$  & $1150$ & $13.1$ & $90.6$     & $82.6$  & $151.6$     & $138.1$ & $80.1$      & $72.9$    & $5.17$ & $2.54$ \\
$15.0$  & $1150$ & $19.6$ & $37.8$     & $37.3$  & $63.1$      & $62.3$  & $33.3$      & $32.9$    & $5.92$ & $2.09$ \\
$20.0$  & $1150$ & $26.1$ & $19.5$     & $21.8$  & $32.6$      & $36.5$  & $17.2$      & $19.3$    & $6.51$ & $1.84$ \\
$40.0$  & $1150$ & $52.2$ & $3.4$      & $6.4$   & $5.7$       & $10.7$  & $3.0$       & $5.6$     & $8.20$ & $1.41$ \\
$2.0$   & $1300$ & $2.3$  & $7864357.0$& $585.3$ & $13153923.0$& $979.0$ & $6945884.8$ & $517.0$   & $3.02$ & $3.02$ \\
$4.0$   & $1300$ & $4.6$  & $3149.0$   & $464.3$ & $5267.1$    & $776.7$ & $2781.3$    & $410.1$   & $3.81$ & $3.81$ \\
$7.0$   & $1300$ & $8.1$  & $407.8$    & $230.2$ & $682.1$     & $385.0$ & $360.2$     & $203.3$   & $4.59$ & $3.55$ \\
$10.0$  & $1300$ & $11.6$ & $118.2$    & $96.8$  & $197.8$     & $161.8$ & $104.4$     & $85.5$    & $5.17$ & $2.75$ \\
$15.0$  & $1300$ & $17.3$ & $42.9$     & $41.9$  & $71.7$      & $70.1$  & $37.9$      & $37.0$    & $5.92$ & $2.22$ \\
$20.0$  & $1300$ & $23.1$ & $23.0$     & $24.0$  & $38.5$      & $40.2$  & $20.3$      & $21.2$    & $6.51$ & $1.94$ \\
$40.0$  & $1300$ & $46.2$ & $4.1$      & $6.9$   & $6.8$       & $11.6$  & $3.6$       & $6.1$     & $8.20$ & $1.47$ \\
\hline
\end{tabular}
\end{table*}

\begin{acknowledgements}
We acknowledge the Austrian Forschungsf\"orderungsgesellschaft FFG
project ``TAPAS4CHEOPS'' P853993, the Austrian Science Fund (FWF)
NFN project S11607-N16, and the FWF project P27256-N27. NVE
acknowledges support by the RFBR grant No. 15-05-00879-a and
16-52-14006 ANF\_a.
\end{acknowledgements}

\bibliographystyle{aa}
\bibliography{../bbl}

\end{document}